\documentclass[aps,prd,10pt,notitlepage,nofootinbib,superscriptaddress,showkeys,showpacs]{revtex4-1}
\usepackage{amsmath,amssymb,amsthm,latexsym}
\usepackage[english]{babel}
\usepackage{graphicx,color}
\usepackage{xspace}
\usepackage{graphicx}
\usepackage{pifont,dsfont}
\usepackage{marvosym}
\usepackage{slashed}
\usepackage{subfigure}
\usepackage[colorlinks=true, linkcolor=green, filecolor=brown, citecolor=green]{hyperref}

\newcommand{\be}{\begin{equation}}
\newcommand{\ee}{\end{equation}}
\newcommand{\bqa}{\begin{eqnarray}}
\newcommand{\eqa}{\end{eqnarray}}
\newcommand{\bea}{\begin{eqnarray}}
\newcommand{\eea}{\end{eqnarray}}

\newcommand{\N}{\mathds{N}}

\theoremstyle{definition}\newtheorem{definition}{Definition}
\theoremstyle{definition}\newtheorem{lemma}{Lemma}
\theoremstyle{definition}\newtheorem{proposition}{Proposition}
\theoremstyle{definition}\newtheorem{corollary}{Corollary}
\theoremstyle{definition}\newtheorem{application}{Application}

\DeclareMathOperator{\tr}{tr}

\DeclareMathOperator{\NCP}{NCP}

\begin{document}

\title{\Large \bf Revisiting random tensor models at large N via the Schwinger-Dyson equations}

\author{{\bf Valentin Bonzom}}\email{vbonzom@perimeterinstitute.ca}
\affiliation{Perimeter Institute for Theoretical Physics, 31 Caroline St. N, ON N2L 2Y5, Waterloo, Canada}

\date{\small\today}

\begin{abstract}
The Schwinger-Dyson Equations (SDEs) of matrix models are known to form (half) a Virasoro algebra and have become a standard tool to solve matrix models. The algebra generated by SDEs in tensor models (for random tensors in a suitable ensemble) is a specific generalization of the Virasoro algebra and it is important to show that these new symmetries determine the physical solutions. We prove this result for random tensors at large N. Compared to matrix models, tensor models have more than a single invariant at each order in the tensor entries and the SDEs make them proliferate. However, the specific combinatorics of the dominant observables allows to restrict to linear SDEs and we show that they determine a unique physical perturbative solution. This gives a new proof that tensor models are Gaussian at large N, with the covariance being the full 2-point function.
\end{abstract}

\medskip

%\noindent  Pacs numbers: 02.10.Ox, 04.60.Gw, 05.40-a

\maketitle

%%%%%%%%%%%%%%%%%%%%%%%
\section*{Introduction}
%%%%%%%%%%%%%%%%%%%%%%%

Matrix models have provided a good description of two-dimensional quantum gravity coupled to matter (equivalently, non-critical strings) \cite{mm-review-difrancesco}. In the scaling limit, they provide access to Liouville gravity coupled to critical (unitary or not) matter with central charge $c<1$. While there exist various ways to solve matrix models, the Schwinger-Dyson Equations (SDEs, also known as loop equations) have become a standard powerful tool. They enable to probe the correlators at all orders in the $1/N$ expansion and the topological expansion \cite{topological-expansion} has been developed as an intrinsic method to solve them. In the double-scaling limit, the SDEs are equivalent to the string equation, which in turn corresponds to some integrable hierarchies (depending on the model) \cite{dijkgraaf-loop-eqs, fukuma-SDEs,martinec,alvarez-gaume, morozov-integrable}.

However the SDEs should not be only considered as a tool, as they probe the key features of matrix models very deeply. In particular, they can be recast in terms of differential operators which generate (half) a Virasoro algebra. This naturally leads to the fermion gas formalism which makes the initially hidden, conformal symmetry manifest, and consequently clarify the relationship of matrix models to 2d conformal field theories \cite{kostov-cft-techniques, banks-macro-loops, ginsparg-moore-lectures}.

It is natural to extend those results to higher dimensions and the proposal was made in the early 90s to use tensors instead of matrices (objects with more indices) \cite{ambjorn-3d-tensors, sasakura-tensors, gross-tensors}. Rank $d$ tensor models indeed generate discrete spacetimes in dimension $d$ and are thus of high interest in different approaches to quantum gravity \cite{tensor-track, carrozza-U1-gft, oriti-gft, quantugeom2}, e.g. (causal) dynamical triangulations \cite{david-revueDT, ambjorn-scaling4D, ambjorn-revueDT, ambjorn-houches94, ambjorn-book} and loop quantum gravity \cite{bahr-finite-groups} (and more generally sit as a natural generalization of matrix models wherever the latter are relevant -- see \cite{p-spin-universality} for an application to disordered systems).

A large $N$ limit has been found only quite recently for tensor models \cite{Gur3, Gur4, GurRiv} , but the subsequent developments have rapidly expanded. The large $N$ contributions are specific discretizations of the $d$-sphere, known as \emph{melonic} \cite{critical-colored}, which provide an analytic description of the universality class of the Branched Polymer (BP) phase of Euclidean dynamical triangulations \cite{ambjorn-d>1, ambjorn-BP} (previously known from numerical simulations). Moreover, the multi-critical behaviors \cite{1tensor} can be interpreted as critical, non-unitary matter \cite{multicritical-dimers, harold-hard-dimers}, just like in 1-matrix models \cite{kazakov-matter}.

Tensor models not only reproduce the statistical properties of the BP phase, but provide a new way to understand dynamical geometries through the SDEs. The algebra they generate was found at large $N$ in \cite{tree-algebra}, and extended at all orders in \cite{bubble-algebra}. Geometrically, while the loop equations of matrix models describe the disc amplitude, the tensor SDEs describe the ball amplitude. The generators are labeled by boundary triangulations.% (equivalently described by equivalence classes of rooted trees in RRR).

As a first step towards a better understanding of this new symmetry algebra, we prove that the SDEs admit a unique physical solution in perturbations. This means that the symmetries completely determine the solution of the model, similarly to the conformal symmetry in two dimensions. This unique physical perturbative solution is obviously the same as found in \cite{universality, 1tensor}, by means of scaling arguments and a precise investigation of the combinatorics of the leading order contributions in the Feynman expansion. Thus, our method has the advantage that it bypasses the Feynman expansion, just like solving the loop equations at large $N$ does not require the knowledge of the planar sector in matrix models.

The main difficulty compared to matrix models is the proliferation of observables. Indeed, at each order in the tensor entries, tensors allow more than a single $U(N)$ invariant (i.e. several boundary triangulations with the same number of simplices) and the SDEs are precisely equations on the expectation values of these observables. However only a specific subset of observables is relevant at large $N$ and our analysis will consider this family as an input. The structure of the relevant observables is such that we can focus on \emph{linear} equations, in contrast with matrix models. This linearity provides an alternative way to \cite{universality} to explain why tensor models are Gaussian at large $N$, the covariance being the full 2-point function\footnote{Notice that this universality theorem does not prevent critical behaviors as the covariance itself is a non-trivial function of the couplings which develops non-analytic singularities \cite{1tensor}.}. Though the sub-leading corrections to expectation values in the $1/N$ expansion are not known, it is clear that non-linearities eventually come into the game. We will restrict here to the large $N$ limit.

The organization is as follows. In the Sec. \ref{sec:review} we briefly review tensor models and the universality theorem. The SDEs are derived in the Sec. \ref{sec:SD}, including our fundamental subset of linear equations. Our main result (the unique physical perturbative solution) is given in the Sec. \ref{sec:generic-case} and we offer a precise comparison with the loop equations of matrix models. The Sec. \ref{sec:14vertices} is an attempt to give a global view on the solutions of the SDEs by comparing the number of observables to the number of independent equations. This is done in a simple example where the relevant observables are shown to be 1-to-1 mapped to non-crossing partitions and we conjecture that an infinity of `initial conditions` is required.

%%%%%%%%%%%%%%%%%%%%%%%
\section{A brief review of tensor models} \label{sec:review}
%%%%%%%%%%%%%%%%%%%%%%%

Building a tensor model requires first a suitable choice of tensor ensemble, defined by its invariance properties (analogously to the Gaussian Unitary Ensemble, or the Gaussian Orthogonal Ensemble in random matrices). A natural choice (the only one for which the large $N$ limit is known to exist) is \emph{an independent unitary invariance on each tensor index}. If $T_{a_1\dotsb a_d}$ are the components of a rank $d$ tensor, $a_i=1,\dotsc,N$ for $i=1,\dotsc,d$, define the following transformation,
\be \label{UN-invariance}
T'_{a_1 \dotsb a_d} = \sum_{b_1,\dotsc,b_d} U^{(1)}_{a_1 b_1} \dotsm U^{(d)}_{a_d b_d}\ T_{b_1 \dotsb b_d},
\ee
where the matrices $U^{(i)}$ are independent unitary matrices of size $N\times N$. The complex conjugated tensor $\bar{T}$ transforms with the complex conjugated matrices. We are interested in functions $f$ over such complex tensors that are invariant under those unitary transformations, $f(T,\bar{T}) = f(T', \bar{T}')$. They are generated by invariant monomials built in the following way \cite{universality}: take $p$ copies of $T$ and $p$ copies of $\bar{T}$ and contract all indices in such a way that an index in the $i$-th position of a $T$ is contracted with an index in the $i$-th position on a $\bar{T}$.

Those invariant monomials are conveniently mapped, in a one-to-one fashion, to $d$-colored bipartite graphs, usually referred to as \emph{bubbles}. Each $T$ is represented by a white vertex and each $\bar{T}$ by a black vertex, giving, say, $p$ black and $p$ white vertices. The indices of each tensor are represented by half-lines labeled by their position, which we call \emph{color}, from 1 to $d$. When two indices are contracted between a $T$ and a $\bar{T}$, they must have the same position hence the corresponding half-lines have the same color and one simply joins them together to form a line labeled with that color. The invariant monomial obtained from a bubble $B$ can be thought as the `trace over the bubble'\footnote{It was often denoted $\tr_B(T,\bar{T})$ in the literature.}, and we denote it $B(T,\bar{T})$.

Bubbles are naturally dual to colored triangulations of $d-1$ pseudo-manifolds. The idea is to associate a $(d-1)$-simplex to each vertex whose half-lines represent the $d$ boundary $(d-2)$-simplices of the $(d-1)$-simplex. Note that the $(d-2)$ simplices inherit the color of the half-lines. Colors further allow to identify all lower-dimensional sub-simplices by considering the sub-bubbles with exactly $k<d$ colors. A bubble line between two vertices describe the gluing of two simplices along a boundary simplex identified by its color. We refer to \cite{1tensor} for details.

Notice that for $d=2$, $T$ is a complex matrix and there is only one bubble with $2p$ vertices: the loop with alternating colors 1 and 2, associated to the trace invariant $\tr(T T^\dagger)^p$. Geometrically, it is dual to a loop with $2p$ lines.

Let $I$ be a finite set, $\{B_i\}_{i\in I}$ a set of bubbles and $\{t_i\}_{i\in I}$ a set of couplings. A generic action for tensor models is
\be \label{action}
S(T,\bar{T}) = T\cdot \bar{T} + \sum_{i\in I} t_i\,B_i(T,\bar{T}),
\ee
where $T\cdot \bar{T} = \sum_{a_i} T_{a_1\dotsb a_d} \bar{T}_{a_1\dotsb a_d}$ is the quadratic part (associated with the bubble formed by two vertices connected together by $d$ lines). The partition function $Z$ and the free energy $F$ are given by
\be
\exp -N^d\,F = Z = \int [dT\,d\bar{T}]\ \exp -N^{d-1}\,S(T,\bar{T}).
\ee
Such integrals are usually\footnote{About constructive aspects and full summability, see \cite{universality}.} understood as power series in the couplings (perturbed Gaussians). It can be shown that $F$ has a $1/N$ expansion which starts at order $\mathcal{O}(1)$, \cite{1tensor}.

\begin{figure}
 \includegraphics[scale=0.45]{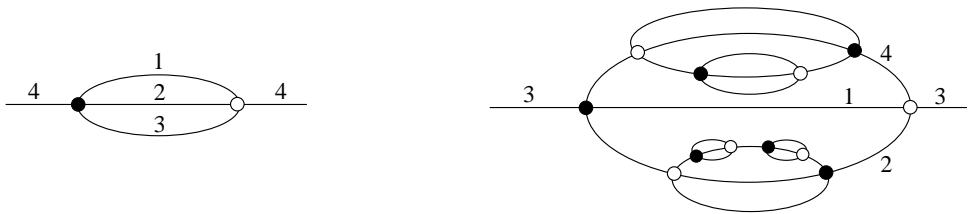}
 \caption{ \label{fig:melon} On the left: an elementary melon with external color 4. On the right: a typical melon on the color 3, with some melonic insertions on the internal lines. To get a melonic graph, one can glue melons next to one another and finally close the two external lines.}
\end{figure}

The natural observables are the bubbles and their expectation values read
\be
\left\langle B(T,\bar{T}) \right\rangle \equiv \frac1{Z} \int [dT\,d\bar{T}]\ B(T,\bar{T})\ \exp -N^{d-1}\,S(T,\bar{T}).
\ee
A distinguished set of bubbles which is of particular importance is the set of \emph{melonic} bubbles.
\begin{definition} \label{def:melon}
The \emph{elementary melon with external color $c$} is defined as the 2-point graph made of two vertices $V, \bar{V}$ connected together by $d-1$ lines (having all colors but $c$), with one external half-line of color $c$ attached to the white vertex $\bar{V}$ and the other, of the same color, attached to the black vertex $V$ (see Fig. \ref{fig:melon}). A \emph{melon} is obtained by inserting recursively elementary melons on any (internal) line between $V$ and $\bar{V}$, starting from the elementary melon itself, as in Fig. \ref{fig:melon}. Melons with the same external color can be joined together so as to get closed connected graphs called \emph{melonic bubbles}.
\end{definition}

Notice that melons in a melonic bubble are precisely the connected, 1-particle-irreducible, 2-point sub-graphs. A melon can also be identified by the two vertices $V, \bar{V}$ on which the two external lines are attached. This gives a \emph{canonical} way to associate to a black vertex $V$ a white vertex, denoted $\bar{V}$.

\medskip

The free energy and the Bubble Expectation Values (BEVs) have Feynman expansions onto connected $(d+1)$-colored graphs (bubbles connected together by propagators which are given a fictitious color). The way tensor model have been solved so far heavily relies on those expansions. The main results have been synthesized in \cite{1tensor}. We recall them briefly.

\begin{enumerate}
%A bubble is melonic if one can build it from the graph with two vertices connected with $d$ lines by inserting recursively on any line pairs of vertices connected by $(d-1)$ lines. The Fig. displays a 2-point melonic patch. It will be useful to define a \emph{melon} as a connected, one-particle-irreducible 2-point subgraph of a melonic bubble. Graphically, a melon has two external lines with the same color, say $c$, that we call the \emph{external color of the melon}. One line is attached to a white vertex and the other to a black vertex. These two vertices are connected by $d-1$ lines with all the colors but $c$ and each of them can carry arbitrary melonic 2-point insertions.
 \item \label{item:melonicbubbles}\emph{Melonic bubbles.} Only melonic bubbles are relevant at large $N$. Consequently one can restrict the sum in \eqref{action} to melonic bubbles only. In the remaining of the paper \emph{all bubbles are considered melonic}. The $1/N$ expansion of their BEVs starts at order $\mathcal{O}(N)$,
\be
\left\langle B(T, \bar{T})\right\rangle = N (K_B + \mathcal{O}(1/N)).
\ee
$K_B$ is the large $N$ amplitude, given by the theorem below. For a non-melonic bubble, $\lim_{N\to \infty} \langle B(T,\bar{T})\rangle /N =0$.

 \item \emph{Gaussian universality.} The model is Gaussian\footnote{A Gaussian theory is defined by the fact that correlators are given as sums over Wick pairings. However in matrix and tensor models, not all pairings are equivalent at large $N$. See the Sec. \ref{sec:gaussian} for details.} at large $N$, with covariance the full 2-point function. In particular the BEVs write
\be \label{universality}
\frac1N\,\left\langle B(T,\bar{T}) \right\rangle = G^{\frac{|B|}{2}},
\ee
where $|B|$ is the number of vertices of $B$ and $G=\langle T.\bar{T}\rangle/N$. All dependence on the coupling constants $t_i$ are carried in $G$. The latter satisfies an algebraic equation (which comes from combining \eqref{universality} with a Schwinger-Dyson equation)
\be \label{algebraicG}
1-G - \sum_{i\in I} p_i\,t_i\,G^{p_i} = 0,
\ee
where $2p_i$ is the number of vertices of $B_i$, and with the condition $G=1$ when all the couplings go to zero.

\end{enumerate}

In this paper, we will take the item \ref{item:melonicbubbles} as granted, as it comes from scaling arguments and amounts to say that we have identified the dominant observables. For $d=2$ it reduces to a quite trivial statement, as it is equivalent to say that the observables are $\langle \tr(MM^\dagger)^p\rangle$ and that their expectation values start like $\mathcal{O}(N)$. In tensor models it however becomes a less trivial assertion. We have not found a way to bypass that argument (i.e. derive the dominance of the melonic bubbles independently) and we actually find it reasonable to start our study with a given set of relevant observables. Then the purpose of the present paper is to find the key equations \eqref{universality} and \eqref{algebraicG} by relying only on Schwinger-Dyson equations and without any use of the Feynman expansion of the BEVS onto $(d+1)$-colored graphs.

In addition to the item \ref{item:melonicbubbles}, we will taken as granted the large $N$ factorization
\be \label{factorization}
\left\langle B(T,\bar{T})\ B'(T,\bar{T}) \right\rangle = \left\langle B(T,\bar{T}) \right\rangle\ \left\langle B'(T,\bar{T}) \right\rangle.
\ee
This is the same assumption that is used in matrix models. It only relies on scaling arguments\footnote{It holds generically in any theory where the Feynman graphs have faces which bring positive powers of $N$. The dominant graphs contributing to the disconnected part of $\left\langle B(T,\bar{T})\ B'(T,\bar{T}) \right\rangle$ have more faces than any connected contribution.}.

%%%%%%%%%%%%%%%%%%%%%%%
\section{The Schwinger-Dyson equations} \label{sec:SD}
%%%%%%%%%%%%%%%%%%%%%%%

The SDEs and their algebra have been presented in \cite{bubble-algebra}. Since this is not quite standard yet, we re-derive them in this section.

%%%%%%%%%%%%%%%%%%%%%%%
\subsection{Bubble insertions}
%%%%%%%%%%%%%%%%%%%%%%%

The simplest SDE is derived from the identity
\be \label{simplestSD}
\frac1Z\ \sum_{a_1,\dotsc,a_d} \int [dT\,d\bar{T}]\ \frac{\partial}{\partial T_{a_1 \dotsb a_d}}\Bigl( T_{a_1\dotsb a_d}\ e^{-N^{d-1}( T\cdot \bar{T} + \sum_{i\in I} t_i\,B_i(T,\bar{T}))}\Bigr) = 0.
\ee
By taking the derivative explicitly and simplifying by $N^d$, one gets
\be \label{firstSD}
1 - \Bigl\langle \frac1N\,T\cdot\bar{T}\Bigr\rangle - \sum_{i\in I} p_i\,t_i\,\Bigl\langle \frac1N\,B_i(T,\bar{T})\Bigr\rangle = 0.
\ee
It is a (linear) equation which relates the BEVs together at all orders of the $1/N$ expansion. It is however not closed and one way to close it at large $N$ is to use of the Gaussian universality property \eqref{universality}, which turns \eqref{firstSD} into an equation on $G$, namely \eqref{algebraicG}.

But we have decided not to use the Gaussian universality (and instead to derive it from the SD equations), which means we have to write other SD equations.
\begin{definition}
Let $V$ be a white vertex in $B$. The \emph{open bubble} $B\smallsetminus V$, with tensor components $(B\smallsetminus V)_{a_1 \dotsb a_d}$, is obtained by removing the vertex $V$ from the bubble (and the corresponding $T_{a_1 \dotsb a_d}$ in the invariant monomial), so that there are $d$ open half-lines carrying the tensor indices $a_1,\dotsc,a_d$. These half-lines hang out from black vertices, hence $B\smallsetminus V$ transforms like a $\bar{T}$.
\end{definition}
%First we define a bubble $B$ open at a white vertex $V$, denoted $B\smallsetminus V$, as the bubble $B$ with the vertex $V$ is simply removed, leaving $d$ open half-lines which carry tensor indices (they are not contracted anymore). This open bubble has tensor components $(B\smallsetminus V)_{a_1\dotsb a_d}$ which transform like $T_{a_1\dotsb a_d}$ (they would transform like $\bar{T}$ if the bubble was opened on a black vertex), because the $d$ open half-lines come from black vertices and are thus indices of tensors $T$.

\begin{figure}
 \includegraphics[scale=0.55]{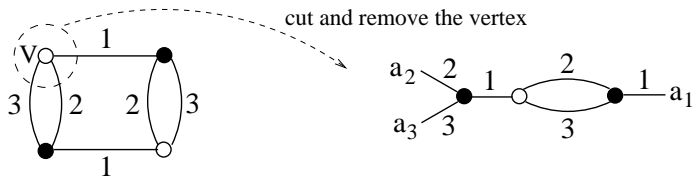}
 \caption{ \label{fig:open-bubble} When the vertex $V$ is cut and removed from the bubble on the left, one obtains a tensorial object, which transforms like a $\bar{T}$. In this example, $d=3$ and $(B\smallsetminus V)_{a_1 a_2 a_3} = \sum_{b_1, b_2, b_3} \bar{T}_{a_1 b_2 b_3} T_{b_1 b_2 b_3} \bar{T}_{ b_1 a_2 a_3}$.}
\end{figure}

An example is given in the Fig. \ref{fig:open-bubble}. One defines similarly the bubbles $B\smallsetminus \bar{V}$ open on a black vertex $\bar{V}$, which transform like a $T$. If $V \in B$ and $\bar{V}'\in B'$, the open bubbles $B\smallsetminus V$, $B'\smallsetminus \bar{V}'$ can be contracted on their free indices to get an invariant under \eqref{UN-invariance}. We denote it
\be
(B\smallsetminus V)\cdot (B'\smallsetminus \bar{V}') \equiv \sum_{a_1,\dotsc,a_d} (B\smallsetminus V)_{a_1 \dotsb a_d} (B'\smallsetminus \bar{V}')_{a_1 \dotsb a_d}.
\ee
This operation has been called \emph{bubble gluing} in \cite{bubble-algebra}.

Open bubbles appear naturally in the derivatives of a bubble invariant,
\be \label{bubble-derivative}
\frac{\partial B(T,\bar{T})}{\partial T_{a_1 \dotsb a_d}} = \sum_{V\in B} (B\smallsetminus V)_{a_1 \dotsb a_d},\qquad \frac{\partial B(T,\bar{T})}{\partial \bar{T}_{a_1 \dotsb a_d}} = \sum_{\bar{V}\in B} (B\smallsetminus \bar{V})_{a_1 \dotsb a_d}.
\ee
What happens when an open bubble is opened a second time? That gives an object with more indices (and possibly disconnected as a graph, i.e. which factorizes as the product of two open bubbles). This can be repeated several times. Conversely, one can take a bubble opened several times and sum over some indices as long as one contracts an index of a $T$ with one of a $\bar{T}$ (to satisfy \eqref{UN-invariance}). This operation has been called \emph{bubble contraction} in \cite{bubble-algebra}.

Open bubbles are covariant objects that can be used as insertions to generalize \eqref{simplestSD}. SDEs are thus labeled by open bubbles and come from the identity
\be
\frac1Z\ \sum_{a_1,\dotsc,a_d} \int [dT\,d\bar{T}]\ \frac{\partial}{\partial T_{a_1 \dotsb a_d}}\Bigl( (B\smallsetminus \bar{V})_{a_1\dotsb a_d}\ e^{-N^{d-1}( T\cdot \bar{T} + \sum_{i\in I} t_i\,B_i(T,\bar{T}))}\Bigr) = 0.
\ee
We have to distinguish three types of contributions.
\begin{enumerate}
 \item The simplest one is when the derivative acts on the quadratic part of the action. Indeed $\partial (T\cdot \bar{T})/\partial T_{a_1 \dotsb a_d} = \bar{T}_{a_1 \dotsb a_d}$ and by definition, $\sum_{a_1,\dotsc,a_d} (B\smallsetminus \bar{V})_{a_1\dotsb a_d} \bar{T}_{a_1 \dotsb a_d} = B(T, \bar{T})$. Therefore
\be \label{middle-term}
-\frac{N^{d-1}}{Z}\ \sum_{a_1,\dotsc,a_d} \int [dT\,d\bar{T}]\ (B\smallsetminus \bar{V})_{a_1\dotsb a_d}\ \frac{\partial\,(T\cdot\bar{T})}{\partial T_{a_1 \dotsb a_d}}\ e^{-N^{d-1}( T\cdot \bar{T} + \sum_{i\in I} t_i\,B_i(T,\bar{T}))} = -N^{d-1}\ \left\langle B(T,\bar{T})\right\rangle.
\ee

 \item Another contribution comes from taking the derivative of the bubble terms of the action. Using \eqref{bubble-derivative}, the derivative of a bubble $B_i$ acts on its $p_i$ white vertices $V_i$ and produces for each of them the open bubble $B_i\smallsetminus V_i$ which transforms like a $\bar{T}$. Hence
\be \label{last-term}
\begin{aligned}
&-\frac{N^{d-1}}{Z}\ \sum_{a_1,\dotsc,a_d} \int [dT\,d\bar{T}]\ (B\smallsetminus \bar{V})_{a_1\dotsb a_d}\ \sum_{i\in I} t_{i}\,\frac{\partial\,B_i(T,\bar{T})}{\partial T_{a_1 \dotsb a_d}}\ e^{-N^{d-1}( T\cdot \bar{T} + \sum_{j\in I} t_j\,B_j(T,\bar{T}))}\\
&= -N^{d-1} \sum_{i\in I} t_{i} \sum_{V_i\in B_i}\ \left\langle \sum_{a_1,\dotsc,a_d} (B\smallsetminus \bar{V})_{a_1\dotsb a_d}\,(B_i\smallsetminus V_i)_{a_1\dotsb a_d} \right\rangle\\ 
&= -N^{d-1} \sum_{i\in I} t_{i} \sum_{V_i\in B_i}\ \Bigl\langle (B\smallsetminus \bar{V})\cdot (B_i\smallsetminus V_i
) \Bigr\rangle.
\end{aligned}
\ee
Here the sum over $V_i$ runs over the white vertices of $B_i$. When $B$ and $B_i$ are melonic, all the resulting bubbles $(B\smallsetminus \bar{V})\cdot (B_i\smallsetminus V_i)$ are melonic, with $|B|+2p_i-2$ vertices, and the scaling is the same as in \eqref{middle-term}.

 \item The last contribution is the derivative of the open bubble $B\smallsetminus \bar{V}$. It produces a sum over its $|B|/2$ white vertices, where for each term the white vertex is removed. That opens the bubble a second time, and the resulting open half-lines carry the indices $a_1,\dotsc,a_d$. They transform like a $\bar{T}$, and performing the sum over the indices connects these half-lines with those of $B\smallsetminus \bar{V}$, producing an invariant which we denote $B\smallsetminus \bar{V}\smallsetminus V'$ (this is a bubble contraction). We can write
\be \label{first-term}
\frac1Z\ \sum_{a_1,\dotsc,a_d} \int [dT\,d\bar{T}]\ \frac{\partial\,(B\smallsetminus \bar{V})_{a_1\dotsb a_d}}{\partial T_{a_1 \dotsb a_d}}\ e^{-N^{d-1}( T\cdot \bar{T} + \sum_{i\in I} t_{i}\,B_i(T,\bar{T}))} = \sum_{V'\in B} \left\langle B\smallsetminus \bar{V}\smallsetminus V'(T,\bar{T}) \right\rangle.
\ee
Each $B\smallsetminus \bar{V}\smallsetminus V'$ is a graph with two vertices less than $B$, and is typically a disconnected set of bubbles (see below).% Depending on the positions of $V$ and $V'$ in $B$, $B\smallsetminus V\smallsetminus V'$ can be a disconnected bubble. We also include in our notation the case where $V'$ is a vertex of $B\smallsetminus V$ which possesses some of the open half-lines. Then one gets a free sum, hence a power of $N$, for each. We can think about these powers of $N$ as trivial graphs with only one closed line and zero vertex.
\end{enumerate}

%%%%%%%%%%%%%%%%%%%%%%%
\subsection{The fundamental large N equations}
%%%%%%%%%%%%%%%%%%%%%%%

One can write an exact equation, which holds at all orders in the $1/N$ expansion, by summing \eqref{first-term}, \eqref{middle-term} and \eqref{last-term} together. But there is a simplification at large $N$, as we have to keep only the melonic contributions in \eqref{first-term} which scale like \eqref{middle-term} and \eqref{last-term}, i.e. $\mathcal{O}(N^d)$. It turns out there is only one such contribution, and this can be proved as follows.

\begin{figure}
 \includegraphics[scale=0.6]{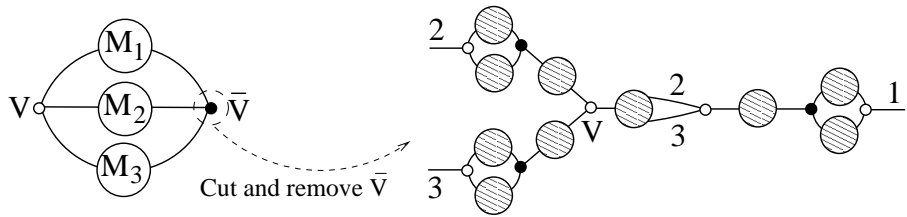}
 \caption{ \label{fig:typical-open-bubble} On the left: the bubble $B$ with the canonical pair $V,\bar{V}$ and 2-point insertions $M_1,\dotsc,M_d$, for $d=3$. On the right: the open bubble $B\smallsetminus \bar{V}$, where the striped circles are melonic insertions (which make the structure of $M_1,\dotsc,M_d$ more explicit). Cutting and removing one white vertex leads to disconnected components. To reach the maximal scaling $N^d$, $d$ disconnected pieces are required. The only way to get them is to cut and remove the vertex $V$ canonically associated to $\bar{V}$.}
\end{figure}

For a melonic $B$, the vertex $\bar{V}$ is part of a canonical pair $(V, \bar{V})$ and one can always draw $B$ like on the left of the Fig. \ref{fig:typical-open-bubble}. The lines of color $1,\dotsc,d$ carry 2-point insertions $M_1,\dotsc, M_d$ that we first assume to be non-trivial. When $\bar{V}$ is removed, one gets the open bubble on the right of the Fig. \ref{fig:typical-open-bubble}, where $M_1,\dotsc,M_d$ have been expanded to describe all typical white vertices. To get $B\smallsetminus \bar{V} \smallsetminus V'$, a white vertex $V'$ has to be removed. Each disconnected piece in $B\smallsetminus \bar{V} \smallsetminus V'$ scales like $N$, so that to reach the scaling $\mathcal{O}(N^d)$, $B\smallsetminus \bar{V} \smallsetminus V'$ must contain $d$ disconnected pieces. It can be checked explicitly using the figure that there is only one way to get $d$ disconnected pieces, which happens for $V'=V$, the vertex canonically associated with $\bar{V}$. The bubbles which result from the contraction are simply the 2-point sub-graphs $M_1,\dotsc,M_d$ which are closed by joining their two external lines (abusing the notation, we still denote the corresponding bubbles $M_c$, for $c=1,\dotsc,d$). %For a melonic $B$, the vertex $\bar{V}$ is part of a canonical pair $(V, \bar{V})$ defining a melon $M_d$ (see the definition \ref{def:melon}), with external color, say, $d$. The open bubble $B\smallsetminus \bar{V}$ can be drawn like in the Fig. without loss of generality. To get $B\smallsetminus \bar{V} \smallsetminus V'$, one has to remove the white vertex $V'$. That opens $d$ lines of all colors that have to be connected with the half-lines of the same color in $B\smallsetminus \bar{V}$.  For simplicity let us first assume that the 2-point insertions on the colors $c=1,\dotsc,d-1$ between $V$ and $\bar{V}$ are non-trivial. They correspond to (not necessarily 1PI) 2-point subgraph $M_i$ with external colors $c=1,\dotsc,d-1$. Each disconnected piece in $B\smallsetminus \bar{V} \smallsetminus V'$ scales like $N$, so that to reach the scaling $\mathcal{O}(N^d)$, $B\smallsetminus \bar{V} \smallsetminus V'$ must contain $d$ disconnected pieces. The white vertices which are displayed in the Fig. \ref{fig:typical-open-bubble} accounts for all possibilities (up to color relabeling) of removing a $V'$. It can be checked explicitly that there is only one way to get $d$ disconnected pieces, which happens for $V'=V$, the vertex canonically associated with $\bar{V}$. The resulting bubbles are simply the 2-point sub-graphs $M_i$ which are closed by joining their two external lines together (abusing the notation, we still denote the corresponding bubble $M_c$), for $c=1,\dotsc,d-1$, plus the bubble $B$ with the melon $M_d$ removed, denoted $B\smallsetminus M_d$. 
Using the large $N$ factorization \eqref{factorization}, we are finally led to
\be \label{SD}
\prod_{c=1}^{d} \left\langle M_c (T,\bar{T})\right\rangle - N^{d-1} \left\langle B(T,\bar{T})\right\rangle - N^{d-1} \sum_{i\in I} t_{i} \sum_{V_i\in B_i}\ \left\langle (B\smallsetminus \bar{V})\cdot (B_i\smallsetminus V_i) \right\rangle =0.
\ee
If some of the $M_c$s are trivial, i.e. they have no vertices and just consist of lines of color $c$, closing them produces loops with no vertices and free sums on the indices $a_c$, each of which simply producing a factor $N$.

The main difficulties with this set of equations are:
\begin{itemize}
 \item the non-linearities, which come from taking the derivative of the open bubble,

 \item the proliferation of observables: from the bubble $B$, the equation generates all the possible gluings of $B\smallsetminus \bar{V}$ with the bubbles $B_i$ contained in the action. This feature contrasts with the loop equations of matrix models by bringing an additional combinatorial difficulty.
\end{itemize}
%We will argue in the present paper that it is possible to restrict attention to well-chosen families of observables such that the non-linearities are avoided (among the 2-point subgraphs $(M_c), B\smallsetminus M_d$, only one is non-trivial) and the bubble proliferation is under control.

%classifying the equations into well-chosen families of equations allows to avoid non-linearities and to control the bubble proliferation.

%%%%%%%%%%%%%%%%%%%%%%%
%\subsection{Two straightforward applications}
%%%%%%%%%%%%%%%%%%%%%%%

The melonic family of bubbles is constructed by recursive insertions of the elementary melon on any line. It turns out that this is the reason why it is going to be sufficient to focus on the SDEs for $B\smallsetminus \bar{V}$ where $\bar{V}$ is the black vertex of an elementary melon. The equations are given in the following lemma.

\begin{lemma} \label{lemma:linear}
\emph{Linear equations for elementary melons.} A melonic bubble $B$ always has at least one elementary melon $M$, with external vertices $V, \bar{V}$, and without loss of generality, external color $d$. The SDE for $B\smallsetminus \bar{V}$ involves the 2-point subgraphs $M_c$, $c=1,\dotsc,d-1$, which are just closed lines with no vertices, hence contributing as $N^{d-1}$. Define the bubble $B\smallsetminus M_d$ as $B$ with the elementary melon $M_d$ replaced by a line of color $d$. Then \eqref{SD} becomes
\be
\left\langle B\smallsetminus M_d \right\rangle - \left\langle B \right\rangle - \sum_{i\in I} t_i \sum_{V_i\in B_i} \left\langle (B\smallsetminus \bar{V})\cdot (B_i\smallsetminus V_i) \right\rangle =0.
\ee
\end{lemma}

These are linear equations which describe the creation/annihilation of an elementary melon $M_d$. The fact that this is a complete set at leading order and that they are linear will quite directly lead to the universal property \eqref{universality}.

%The fact that some SD equations are linear is no more than a curiosity, unless it can be proved that they can be organized into closed sets of equations. This will actually be done in the Section ???. The fact that the set of linear equations is sufficient to solve the model is due to the fact that they are associated to removing/inserting an elementary melon and that any melonic bubble can be obtained from any other melonic bubble by a sequence of removals/insertions of elementary melons.

\medskip

An interesting, direct application of the above Lemma is given below (but it will not be used in the proof of the universality result).

\begin{application} \label{lemma2}
\emph{Equalities between sums of BEVs.} Let $G$ be a 2-point (melonic) graph, say, with external color 1. Then,
\be \label{eq:lemma2}
\sum_{i\in I} t_i \sum_{V_i\in B_i} \left\langle \begin{array}{c} \includegraphics[scale=0.5]{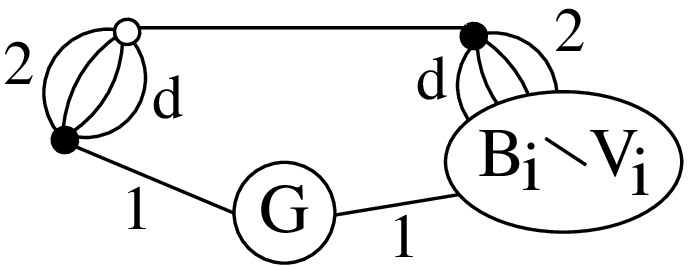} \end{array} \right\rangle = \sum_{i\in I} t_i \sum_{V_i\in B_i}  \left\langle \begin{array}{c} \includegraphics[scale=0.5]{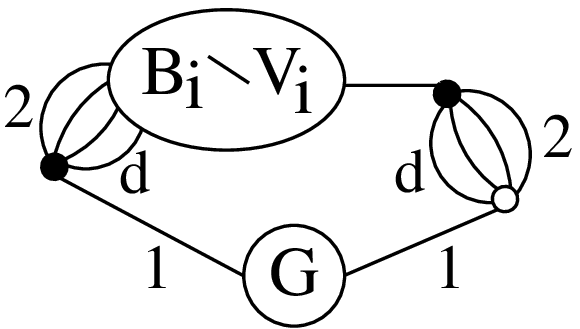} \end{array} \right\rangle.
\ee
\end{application}

The proof goes as follows. Consider the bubble $B$ obtained by closing the graph $G$ with two additional elementary melons next to each other on the color 1. We call $V$ and $V'$ the two white vertices of these melons, and compare the SDEs for $B\smallsetminus V$ and $B\smallsetminus V'$. Both contain the expectation value of $B$. Since in the two cases $B$ is opened on elementary melons, the lemma \ref{lemma:linear} tells us that the SDEs generate a single bubble with two less vertices. This is the same bubble in both cases: $B$ with one elementary melon replaced by a line of color 1. Therefore, the only difference comes from the gluings of $B$ with the bubbles $B_i$ of the action. The proof ends by substracting one equation to the other.

This application becomes more interesting in special cases. For instance if the bubbles in the action have the same number of vertices, then the equation compares the BEVs of bubbles which all have the same number of vertices. It is even better for instance if there are some symmetries on the bubbles which reduce the number of terms of the equation.

%%%%%%%%%%%%%%%%%%%%%%%
\section{The generic 1-tensor model} \label{sec:generic-case}
%%%%%%%%%%%%%%%%%%%%%%%

%%%%%%%%%%%%%%%%%%%%%%%
\subsection{The Gaussian model} \label{sec:gaussian}
%%%%%%%%%%%%%%%%%%%%%%%

Before analyzing the generic model, it is useful to understand the behavior of the Gaussian one. For a covariance $g$, the action is
\be
S(T,\bar{T}) = \frac1{g}\ T\cdot \bar{T}.
\ee
The SDEs of the Lemma \ref{lemma:linear} are particularly simple,
\be
\left\langle B \right\rangle = g\ \left\langle B\smallsetminus M \right\rangle,
\ee
for any elementary melon $M$ in $B$. As any melonic bubble comes from inserting an elementary melon on a smaller bubble, we get with the initial condition $\langle 1 \rangle =1$ that
\be \label{gaussianBEV}
\langle B \rangle = N\,g^p,
\ee
where $p$ is the half number of vertices. This result has a combinatorial interpretation. Indeed a Gaussian measure is standardly defined by the fact that the expectation values are sums over Wick pairings weighted by the covariance. However, in matrix and tensor models not all Wick contractions have the same scaling with $N$, which means that some are suppressed at large $N$. As well-known, in the Gaussian matrix model, only planar contractions survive at large $N$ (and are counted by the Catalan numbers). For the Gaussian tensor model, the BEV \eqref{gaussianBEV} shows that there is a \emph{single} Wick pairing which survives the large $N$ limit. It has been shown to be the pairing which connects the canonical pairs $(T,\bar{T})$ of the melons in $B$ \footnote{In other words, there is a single way to build a melonic graph with $(d+1)$ colors from a melonic graph with $d$ colors and this is the only way to maximize the number of faces. This is the combinatorial reasoning used in \cite{universality}.}.

%%%%%%%%%%%%%%%%%%%%%%%
\subsection{Solving the generic 1-tensor model}
%%%%%%%%%%%%%%%%%%%%%%%

%Now that the 1-bubble model has been solved for an arbitrary bubble, we are going to solve the generic case by an induction on the number of bubbles in the action.
In this section $I$ is a finite set, $\{B_i\}_{i\in I}$ a set of bubbles with associated couplings $\{t_i\}_{i\in I}$. We denote the half-number of vertices of each bubble by $p_i = |B_i|/2$. We will need the following Proposition.

\begin{proposition} \label{lemma:bubble-evaluation}
Let $(F_n(\{t_i\}))_{n\in \N}$ be a sequence of functions such that each $F_n$ has a power series expansion in each $t_i$. Let $E$ be a map which associates to each bubble $B$ a series $E_B(\{t_i\})$ in the couplings $\{t_i\}$ such that the empty bubble is mapped to the constant 1. Assume that for any vertex $V$ of an elementary melon $M$ in a bubble $B$ with $2p$ vertices, $E$ satisfies the equation
\be \label{SDevaluation}
E_{B\smallsetminus M} - E_B - \sum_{i\in I} t_i \sum_{\bar{V}'\in B_i} E_{(B\smallsetminus V)\cdot (B_i\smallsetminus \bar{V}')} = F_p.
\ee
Then the evaluation $E$ on two bubbles with the same number of vertices gives the same function, i.e. $E$ only depends on the number of vertices of the bubbles $B$ and not their specific structure.
\end{proposition}

{\bf Proof.} We proceed by induction on the number of elements in $I$.
\begin{itemize}
 \item When $I$ is empty, $E_{B\smallsetminus M} - E_B$ only depends on the number of vertices of $B$, hence the result follows from a trivial recursion on the number of vertices of the bubbles (as any bubble can be obtained by inserting an elementary melon on a smaller bubble).
 \item Assume the lemma holds for a particular set $I$ and consider an additional bubble $B_0$, with $2p_0$ vertices and coupling $t_0$. We expand $E_B$ as
     \be \label{E_B}
     E_B(t_0,\{t_i\}) = \sum_{n\geq 0} t_0^n\ E_B^{(n)}(\{t_i\}).
     \ee
     The function $F_p$ has a similar expansion as $F_p = \sum_n t_0^n F_p^{(n)}$. The equation satisfied by $E$ becomes at order $n\geq 1$
     \be \label{Fnp}
     E_{B\smallsetminus M}^{(n)} - E_B^{(n)} - \sum_{i\in I} t_i \sum_{\bar{V}'\in B_i} E_{(B\smallsetminus V)\star (B_i\smallsetminus \bar{V}')}^{(n)} = F_p^{(n)} + \sum_{\bar{V}'\in B_0} E_{(B\smallsetminus V)\star (B_0\smallsetminus \bar{V}')}^{(n-1)},
     \ee
     and
     \be \label{F0p}
     E_{B\smallsetminus M}^{(0)} - E_B^{(0)} - \sum_{i\in I} t_i \sum_{\bar{V}'\in B_i} E_{(B\smallsetminus V)\star (B_i\smallsetminus \bar{V}')}^{(0)} = F_p^{(0)}.
     \ee
     At this stage, we would like to apply the lemma for the set $I$ as our induction hypothesis. However the right hand side of \eqref{Fnp} is {\it a priori} not a sequence since it seems that it depends on gluings of bubbles. However, we now show by induction on the order of the perturbation that it only depends on the number of vertices of $B$ and that the lemma for the set $I$ can be applied to \eqref{Fnp}.
     \begin{itemize}
     \item At $n=0$, the lemma can be clearly applied on \eqref{F0p} with the set $I$. As a result $(E_B^{(0)}(\{t_i\})_B$ is actually a sequence of functions since its sole dependence on $B$ is through the number of vertices.
     \item Assume this holds at order $n-1$, for $n\geq 1$, we apply the lemma to \eqref{Fnp} for the set $I$ and find that the dependence of $E^{(n)}_B$ on $B$ is just on the number of vertices of $B$.
     \end{itemize}
     Therefore, the full expansion \eqref{E_B} of $E_B$ only probes bubbles through their number of vertices, which proves the desired property for the set $\{B_i\}_{i\in I}$ supplemented with $B_0$.
\end{itemize}
\qed

%With the same notions as above, we assume that the universal property \eqref{universality} has been proved for the set $\{B_i\}_{i\in I}$. We denote the half-number of vertices of each bubble by $p_i = |B_i|/2$, for $i\in I$. Let $B_0$ be another bubble, with $2p_0$ vertices, and consider the action
We now apply the above proposition to the tensor model with the generic action
\be \nonumber
S(T,\bar{T}) = T\cdot \bar{T} + \sum_{i\in I} t_i\,\tr_{B_i}(T,\bar{T}).
\ee
One sets $E_B = \langle B(T,\bar{T}) \rangle$, $F=0$ and the large $N$ BEVs satify the equation \eqref{SDevaluation} (the Lemma \ref{lemma:linear}).

\begin{corollary}
The bubble dependence of the BEVs is just through their number of vertices. For any bubble with $2p$ vertices, denote $\langle B(T,\bar{T}) \rangle = N\, G_p$. Then the linear SD equations reduce to
\be
G_n - G_{n+1} - \sum_{i\in I} p_i\,t_i\,G_{n+p_i} = 0.
\ee
\end{corollary}

Let us re-organize the set of bubbles $\{B_i\}_{i\in I}$ according to the number of vertices of the bubbles. Set $I=\cup_{p\in P} I_p$, where $P$ is a finite set of integers greater or equal to 2, such that $\{B_i\}_{i\in I_p}$ is the subset which contains the bubbles with $2p$ vertices. We define the coupling at $2p$ vertices as $t_p = \sum_{i\in I_p} t_i$. Then the recursion reads
\be \label{skewed-recursion}
G_n - G_{n+1} - \sum_{p\in P} p\,t_p\,G_{n+p} = 0.
\ee
This linear recursion can be solved in two (equivalent) ways.

\begin{enumerate}
 \item \emph{Solving the recursion.} The characteristic polynomial is
\be \label{char-pol}
\sum_{p\in P} p\,t_p\, X^p +X-1 = p_*\,t_{p_*}\,(X-G)  \prod_\alpha (X-G_{(\alpha)}).
\ee
Here $p_* = \operatorname{Sup} P$ is the maximal number of vertices among the bubbles in the action. When the couplings $\{t_p\}$ go to zero, there is a single root $X=1$. We have denoted $G$ the root which goes to 1 when the couplings go to 0 and we have isolated it on purpose as it is the \emph{physical} root. Assuming that the $P$ roots $G,(G_{(\alpha)})$ of this polynomial are distinct, it comes for any $B$ with $2p$ vertices
\be
\Bigl\langle \frac1N B(T,\bar{T}) \Bigr\rangle = c\,G^p + \sum_\alpha c_\alpha\ G_{(\alpha)}^p,
\ee
for some constants $c$ and $c_\alpha$. The physical interpretation is clear from the analysis of the Gaussian model in the Sec. \ref{sec:gaussian}: each geometric contribution $G_{(\alpha)}^p$, $G^p$, is a large $N$, Gaussian channel with covariance $G_{(\alpha)}, G$. But only one of them goes to 1 when the couplings go to 0 and it is the physical Gaussian channel with covariance $G$. Therefore $\langle B \rangle =N\, G^n$, and $G$ is the full 2-point function. Note that this argument avoids the discussion of the initial conditions which are in principle necessary to solve the recursion.

 \item \emph{With the resolvent.} Natural objects of matrix models are the resolvent and its associated eigenvalue density. While tensors have neither natural multiplication like matrices, nor eigenvalues, the fact that the large $N$ BEVs only depend on an integer (the number of vertices) enables to define a natural analog to the resolvent
\be
\omega(z) \equiv \sum_{p\geq 0} z^{-p-1}\ G_p,
\ee
In matrix models it turns the SDEs into an algebraic equation on $\omega(z)$. We can proceed similarly here, to get
\be
\omega(z) = \frac{G_0+ \sum_{p=2}^{p_*} p\,t_p\,G_{p-1} + \sum_{k=1}^{p_*-1} z^k \sum_{p=k+1}^{p_*} p\,t_p\,G_{p-1-k}}{\sum_{p=2}^{p_*} p\,t_p\,z^p +z - 1},
\ee
where $G_k$ for $k=0,\dotsc,p_*-1$ are the required $p_*$ initial conditions. To avoid setting the initial conditions by hand, we need to identify the physical Gaussian of covariance $G$ (the one which goes to 1 when the couplings go to zero) in the language of the resolvent. This is done by analyzing the structure of its singularities.

We observe that the denominator is the characteristic polynomial \eqref{char-pol}, whose roots are $G, (G_{(\alpha)})$, hence
\be
\omega(z)= \frac{G_0+ \sum_{p=2}^{p_*} p\,t_p\,G_{p-1} + \sum_{k=1}^{p_*-1} z^k \sum_{p=k+1}^{p_*} p\,t_p\,G_{p-1-k}}{p_*\,t_{p_*}\,(z-G) \prod_\alpha (z-G_{(\alpha)})}.
\ee
Since the numerator is a polynomial of degree $p_*-1$, the resolvent has at most $p_*$ poles, $z=G, z=G_{(\alpha)}$, and at least one pole. As only $G$ is physically relevant, we make a \emph{1-pole hypothesis}: the polynomial in the numerator must remove the $p_*-1$ non-physical singularities located on $(G_{(\alpha)})$. This fixes the initial conditions $G_k$, $k=0,\dotsc,p_*-1$, up to a global scale. The latter is fixed by the trivial condition $G_0=\langle 1 \rangle =1$. This reasoning leads to the final form of the resolvent
\be
\omega(z) = \frac1{z-G}.
\ee
The associated `eigenvalue distribution' $\rho(\lambda)$ is obtained as usual by taking the discontinuity of the resolvent across the real line,
\be
\rho(\lambda) = \delta(\lambda-G).
\ee
%which is quite obvious since there is no Vandermonde contribution at large $N$ to prevent the eigenvalues from falling into the potential well at $V'(G)=0$. Also trivially, $\langle \frac1N \tr(TT^\dagger)^n\rangle = \int d\lambda\,\rho(\lambda)\,\lambda^n = G^n$.
\end{enumerate}

\subsection{Comparison with matrix models} \label{sec:skewed}
%%%%%%%%%%%%%%%%%%%%%%%

The above resolvent and `eigenvalue distribution' describe a system of non-interacting particles, all falling in the same potential well at $z=G$. This is in contrast with matrix models, but it is actually what would happen in a matrix model if the famous Vandermonde contribution, which acts like a Coulomb gas repulsion between the eigenvalues, could be removed. Such a matrix model has actually been constructed in \cite{toy-double-scaling}. Further, the universality property \eqref{universality} ensures that all physical quantities of the generic 1-tensor model can be evaluated with this model.

\begin{figure}
 \includegraphics[scale=0.45]{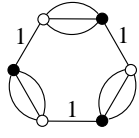}
 \caption{ \label{fig:skewed-loop} Such bubbles can be re-written as loops observables in a matrix form $\tr(TT^\dagger)^p$, with $p=3$ in the picture, for a matrix $T_{aA}$ where $a=1,\dotsc,N$ is the index on the color 1, and $A$ ranging from 1 to $N^{d-1}$ describes all the colors $2,\dotsc,d$, with $d=4$ here.}
\end{figure}

It is a matrix model for `very rectangular' matrices, of size $N\times N^{d-1}$, that is built as a tensor model with very specific bubbles. The bubbles entering the action are chosen such that the indices $a_2,\dotsc,a_d$ of $T_{a_1 a_2 \dotsb a_d}$ are always contracted all at the same time with the corresponding indices of a single $\bar{T}$, as in the Fig. \ref{fig:skewed-loop}. The associated bubbles look like loops with some lines of color 1 and the others being `fat lines' with colors $2,\dotsc,d$. These invariants can be written in a matrix form. Introduce a `fat index' of size $N^{d-1}$, denoted by a capital letter, $A=(a_2,\dotsc,a_d)$ and write the tensor like a matrix $(T_{aA})$ of size $N\times N^{d-1}$ and its complex conjugate $\bar{T}_{aA} = (T^\dagger)_{Aa}$. The corresponding action is
\be
S(T,T^\dagger) = \tr(TT^\dagger) + \sum_{p=2}^{p_*} t_p\,\tr(TT^\dagger)^p.
\ee
This model was analyzed in details in \cite{toy-double-scaling}. Let us reproduce briefly the points that are most relevant to us. Upon introducing the eigenvalues $(\lambda_i)_{i=1,\dotsc,N}$ of $TT^\dagger$, the partition function reads
\be
Z = \int \prod_{i=1}^N d\lambda_i \ \exp -N^{d-1} \left(\sum_{i=1}^N V(\lambda_i) + \frac{N}{N^{d-1}} \biggl( \sum_{i=1}^N \ln\lambda_i -  \frac{2}{N} \sum_{i<j} \ln |\lambda_i-\lambda_j|\biggr)\right),
\ee
with
\be
V(\lambda) = \lambda + \sum_{p=2}^{p_*} t_p\,\lambda^p - \ln \lambda.
\ee
The two $\ln\lambda$ terms are due to the fact we work with a rectangular matrix (they cancel each other for a square matrix). One recognizes $\frac1{N}\sum_{i<j} \ln |\lambda_i-\lambda_j|$ as the famous Vandermonde contribution. It is of order $N$ so when $d=2$ it balances the potential $\sum_i V(\lambda_i)$. It is the usual matrix model philosophy: the Vandermonde term acts like a 2d Coulomb repelling which prevents the eigenvalues to all fall in the minimum of the potential and hence spreads the eigenvalue distribution. However in our case, the Vandermonde is rescaled by $N/N^{d-1}$ and is therefore suppressed as soon as $d>2$. Consequently the saddle point equation decouples the eigenvalues and reads
\be
V'(\lambda) = \frac1{\lambda} \Bigl( \sum_{p=2}^{p_*} p\,t_p\,\lambda^p +\lambda - 1\Bigr) = 0.
\ee
One recognizes the characteristic polynomial \eqref{char-pol} of the BEV recursion. Thus, the large $N$ Gaussian channels with covariance $G, (G_{(\alpha)})$ identified in the tensor model analysis naturally appear here as the solutions of the saddle point approximation.
%Since $G=\frac1{N}\langle \tr TT^\dagger\rangle = \langle \lambda\rangle$, it is indeed a special case of the equation \eqref{algebraicG}.

Let us now show how the loop equations describe the above phenomenon. We start with the identity
\be
\int [dT\,dT^\dagger]\ \frac{\partial}{\partial T_{aA}} \biggl( \Bigl[ \bigl(TT^\dagger\bigr)^n T\Bigr]_{aA}\ \exp -N^{d-1} S(T,T^\dagger)\biggr) = 0.
\ee
Next we evaluate explicitly the derivatives. Using the large $N$ factorization $\langle \tr(TT^\dagger)^k\,\tr(TT^\dagger)^p\rangle = \langle \tr(TT^\dagger)^k\rangle \langle\tr(TT^\dagger)^p\rangle$, one gets
\be
\Bigl\langle \frac1N \tr(TT^\dagger)^n \Bigr\rangle - \Bigl\langle \frac1N \tr(TT^\dagger)^{n+1} \Bigr\rangle - \sum_{p=2}^P p\,t_p\,\Bigl\langle \frac1N \tr(TT^\dagger)^{n+p} \Bigr\rangle
+ \frac1{N^{d-2}}\,\sum_{k=0}^{n-1} \Bigl\langle \frac1N \tr(TT^\dagger)^k \Bigr\rangle\,\Bigl\langle \frac1N \tr(TT^\dagger)^{n-k} \Bigr\rangle = 0.
\ee
All quantities into brackets $\langle\ \rangle$ are of order $\mathcal{O}(1)$, hence for $d>2$ the non-linear terms are suppressed. In other words, the presence of the Vandermonde contribution to the saddle point equation translates into non-linearities in the SD equations. For $d>2$ the natural scaling of tensor models removes those non-linearities and the linear equations of the Lemma \ref{lemma:linear} are recovered.

%%%%%%%%%%%%%%%%%%%%%%%
\section{About observables and other solutions of the Schwinger-Dyson equations} \label{sec:14vertices}
%%%%%%%%%%%%%%%%%%%%%%%

We proved in the previous section that there is a unique physical perturbative solution to the SDEs at large $N$. However, nothing has been said about the full set of solutions to the equations. We propose some preliminary analysis in this Section, based on a specific, simple example, and conjecture that the set of solutions is determined by an infinity of `initial' conditions.

We consider two bubbles $B_1, B_2$, both with four vertices but different color labels, and two different coupling constants $t_1,t_2$,
\be
S(T, \bar{T}) = T\cdot \bar{T} + t_1\ \underbrace{\begin{array}{c} \includegraphics[scale=0.35]{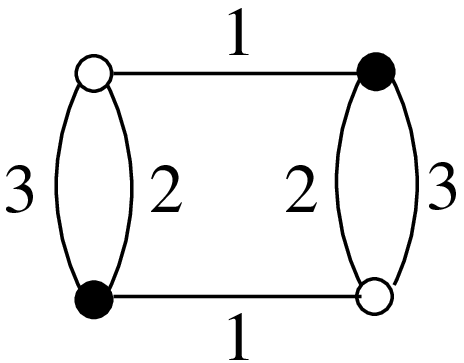} \end{array}}_{B_1(T,\bar{T})} + t_2\ \underbrace{\begin{array}{c} \includegraphics[scale=0.35]{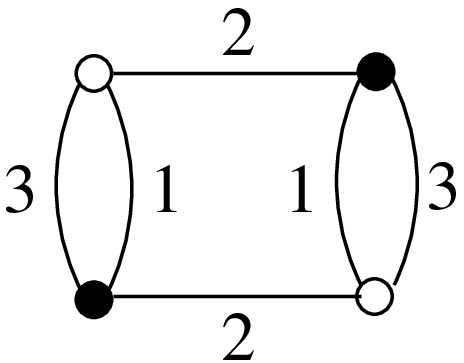} \end{array}}_{B_2(T,\bar{T})}
\ee
Note that the bubbles above have been drawn for $d=3$. To get them for generic $d>3$, it is sufficient to add lines with colors $4,\dotsc,d$ whenever there is already a line of color 3, i.e. between all canonical pairs $(V, \bar{V})$. The model gives a special role to the colors 1 and 2. As a consequence, it is natural to define a specific sub-set of melonic bubbles.
\begin{definition}
 A bubble is said to have \emph{melons on the colors 1 and 2 only} if it can be built from the 2-vertex bubble by recursive insertions of elementary melons on the colors 1 and 2 only. Let $\mathcal{B}_{12}(p)$ be the set of such bubbles with exactly $2p$ vertices, $|\mathcal{B}_{12}(p)|$ the corresponding number of bubbles, and $\mathcal{B}_{12} = \cup_{p\in\N} \mathcal{B}_{12}(p)$.
\end{definition}

To write the SDEs, we need to know the open bubbles obtained from $B_1$ and $B_2$. The two black vertices of $B_1$ (and $B_2$) give the same open bubble $B_1\smallsetminus \bar{V}_1$ (and $B_1\smallsetminus \bar{V}_1$), depicted in Fig. \ref{fig:open-bubble}. Therefore, for any choice of $B$, the SDE generates two larger graphs, one with an additional elementary melon on a line of color 1 on $B$ and the other with an additional elementary melon on a line of color 2. This implies that if one writes a SDE for an open bubble $B\smallsetminus V$ which has melons of external colors 1 and 2 only, the equation only generates other bubbles in $\mathcal{B}_{12}$. We can write the SDEs graphically,
\be
\frac1{N} \left\langle \begin{array}{c} \includegraphics[scale=0.5]{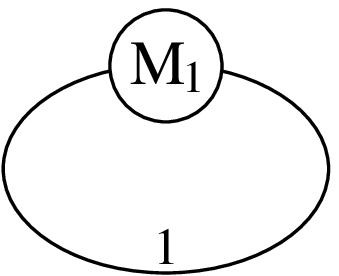} \end{array} \right\rangle \left\langle \begin{array}{c} \includegraphics[scale=0.5]{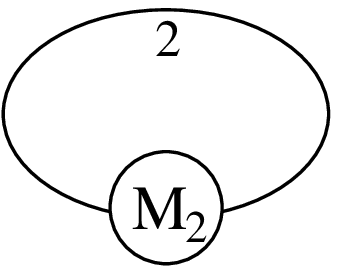} \end{array} \right\rangle - \left\langle \begin{array}{c} \includegraphics[scale=0.5]{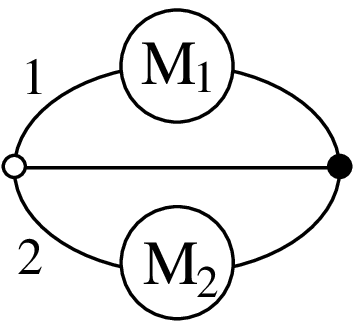} \end{array} \right\rangle - 2t_1 \left\langle \begin{array}{c} \includegraphics[scale=0.5]{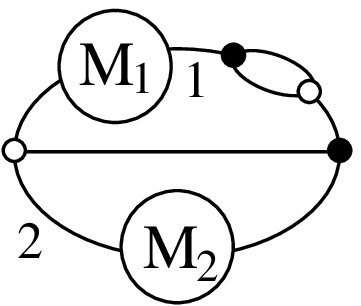} \end{array} \right\rangle -2t_2 \left\langle \begin{array}{c} \includegraphics[scale=0.5]{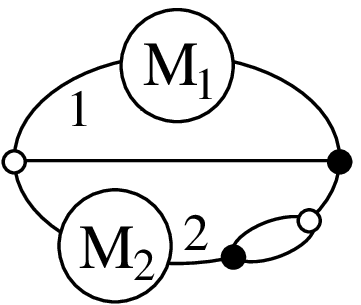} \end{array} \right\rangle = 0.
\ee
Quite clearly, these equations actually generate the whole set $\mathcal{B}_{12}$.

% For $B$ the bubble with only vertices connected by $d$ lines, these graphs are exactly $B_1$ and $B_2$, For $B=B_1$, we get a bubble with 3 melons on the color 1 and another with two melons on the color 1 and one on the color 2 (and symmetrically for $B_2$). By induction, it is clear that all possible melonic bubbles with melons on the colors 1 and 2 only are generated, i.e. all elements of $\mathcal{B}_{12}$. Moreover, any other SD equation involves bubbles with at least one melon on another color.

%%%%%%%%%%%%%%%%
%\subsection{Number of observables versus number of independent SD equations}
%%%%%%%%%%%%%%%%
We will first count the number of observables generated by $B_1$ and $B_2$ from the trivial bubble, and then compare with the number of independent SDEs, for graphs with up to fourteen vertices.

%%%%%%%%%%%%%%%%
\subsection{Melonic bubbles and non-crossing partitions}
%%%%%%%%%%%%%%%%

As mentioned above, the sets $\mathcal{B}_{12}(p)$ are actually independent of $d\geq 3$. If $d>3$, then for any line of color 3 between two canonically associated vertices $V, \bar{V}$, there are also lines of colors $4,\dotsc,d$ between $V$ and $\bar{V}$, and this situation accounts for all lines of colors $3,\dotsc,d$. By removing the lines of colors $4,\dotsc,d$, one obtains an element of $\mathcal{B}_{12}$ for $d=3$. Reciprocally, given an element of $\mathcal{B}_{12}$ for $d=3$, one can add lines of colors $4,\dotsc,d$ between all pairs $(V, \bar{V})$ to go back to the generic case. Therefore we focus on $d=3$ in the following.

One can represent an element of $\mathcal{B}_{12}(p)$ as a bipartite, planar contraction among $2p$ elements. Indeed, one first places the $2p$ vertices of the bubble on a circle, alternating black and white vertices. Using the clockwise convention, we put the color 1 on the $p$ arcs which go from a black vertex to a white vertex, and the color 2 on the $p$ other arcs. Each black vertex must then be connected to a white vertex by a line of color 3, the pairings being allowed only if they result in a planar graph.

\begin{figure}
 \includegraphics[scale=0.75]{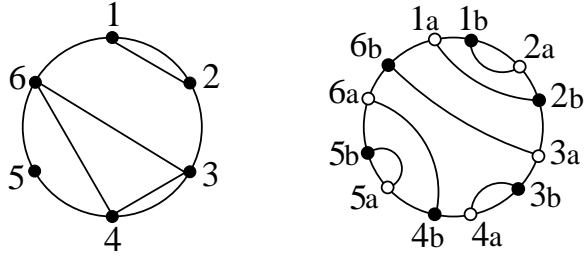}
 \caption{ \label{fig:NCP-melon} One the left: a NCP on 6 elements with labeled vertices. On the right: an element of $\mathcal{B}^{\rm lab}_{12}(p=6)$. A melonic bubble is obtained by putting back the color 2 on each arc between $n_a$ and $n_b$, and the color 1 on the arcs between $n_b$ and $(n+1)_a$, and removing the vertex labels.}
\end{figure}

The vertices of the bubbles in $\mathcal{B}_{12}$ are not labeled. We introduce $\mathcal{B}_{12}^{\rm lab}$ as the set of bubbles whose melons have external colors 1 and 2 only, but now with labeled vertices. We label them $(1_a,1_b,\dotsc,p_a,p_b)$ going clockwise around the circle, where the vertices $(n_a)$ are the white ones and $(n_b)$ the black ones, as shown on the right of the Fig. \ref{fig:NCP-melon}. Then $\mathcal{B}_{12}$ can be described as the equivalence classes of these labeled planar contractions under the action of the rotations $(n_a\mapsto (n+1)_a, n_b\mapsto (n+1)_b)$ (with periodic boundary conditions), and reflections.

Let us now introduce the Non-Crossing Partitions (NCPs) of $[p]=\{1,\dotsc,p\}$ in a graphical way. A partition $\pi$ can be pictured by putting the elements of $[p]$ on a circle, going from 1 to $p$ clockwise, and adding links between the elements of each sub-set as follows. If $n,m$ form a sub-set $\{n,m\}$ with two elements, a line is drawn to connect them. When a sub-set has at least three elements $\{k,l,m,\dotsc\}$, we draw the unique convex polygon whose vertices are $k,l,m,\dotsc$. Finally singlets $\{n\}$ are identified as isolated vertices on the circle. The set $\NCP(p)$ of non-crossing partitions is the set of partitions $\pi$ of $[p]$ with no crossing inside the circle. An example is displayed on the left of the Fig. \ref{fig:NCP-melon}.

\begin{figure}
 \includegraphics[scale=0.85]{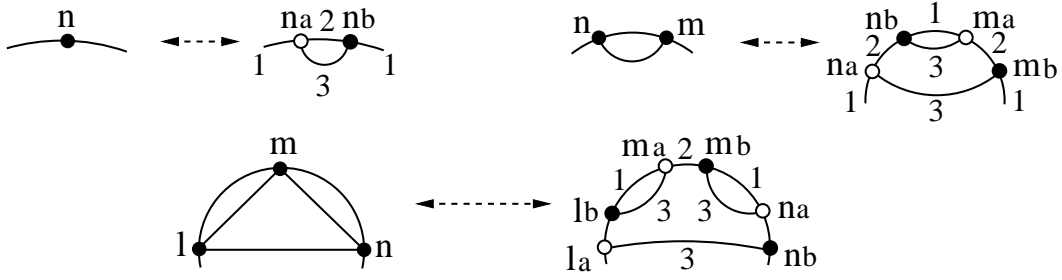}
 \caption{ \label{fig:NCP-melon-map} The rules to map non-crossing partitions to $\mathcal{B}_{12}^{\rm lab}$ and back.}
\end{figure}

We will describe a bijection between $\mathcal{B}_{12}^{\rm lab}(p)$ and $\NCP(p)$ and then mod out rotations and reflections. First we present the map from $\NCP(p)$ to $\mathcal{B}_{12}^{\rm lab}(p)$. We start by splitting each vertex $n=1,\dotsc,p$ into two, denoted $n_a,n_b$, clockwise ordered, and label the arcs between $n_b$ and $(n+1)_a$ with the color 1 and the arcs between $n_a$ and $n_b$ with the color 2. We can distinguish three types of elements in $\pi \in \NCP(p)$:
\begin{itemize}
 \item for each singlet $\{n\} \in \pi$, we draw a line (with color 3) between $n_a$ and $n_b$ (this creates an elementary melon on the color 1),
 \item for $\{n,m\} \in \pi$ with exactly two elements, we draw a line between $n_a$ and $m_b$, and a second line between $n_b$ and $m_a$ (notice that they do not cross),
 \item for a polygon corresponding to $\{n_1,n_2,n_3,\dotsc\} \in \pi$, such that the links are between $n_k$ and $n_{k+1}$, we draw lines from $(n_k)_b$ to $(n_{k+1})_a$.
\end{itemize}
To get the inverse map, one simply reverses the above three rules, which account for all possible patterns inside the bubbles. The correspondance is detailed graphically in the Fig. \ref{fig:NCP-melon-map}. Then, it is easy to see using this map that the rotations $(n_a\mapsto (n+1)_a, n_b\mapsto (n+1)_b)$ are mapped to usual rotations $n\mapsto n+1$ on $\NCP(p)$, and reflections to reflections. As a consequence, the number of bubbles $|\mathcal{B}_{12}(p)|$ is the number of non-crossing partitions of $[p]$ up to rotations and reflections (i.e. dihedral classes). They have been studied in \cite{NCPRR} and are known as \htmladdnormallink{A111275}{http://oeis.org/A111275}$(p)$. The first values, from $p=1$ up to $p=7$, are $1, 2, 3, 6, 10, 24, 49$.

%%%%%%%%%%%%%%%%%
\subsection{Independent SD equations}
%%%%%%%%%%%%%%%%%

Now that we have a good control on the family $\mathcal{B}_{12}$, we come back to the SDEs. The key question is: do the SDEs determine the BEVs up to a finite number of initial conditions? While we have not been able to give a rigorous answer,  we conjecture that the answer is no. It is based on the explicit analysis of the equations on $\mathcal{B}_{12}(p)$ up to $p=7$.

Each SDE relates the BEVs of two bubbles with $2(p+2)$ vertices to the BEVs of bubbles which have less vertices. From the point of view of the two bubbles in $\mathcal{B}_{12}(p+2)$, the equations form a linear system. We will be interested in its rank, to compare it in particular with the number of bubbles $|\mathcal{B}_{12}(p+2)|$. Our conjecture is that the rank of the system is precisely $|\mathcal{B}_{12}(p+2)|-1$, for all $p\geq 0$. To support it, we first write down the equations for a few values of $p$. For $p=0$,
\be \label{p=0}
2t_1 \left\langle \begin{array}{c} \includegraphics[scale=0.35]{B1.eps} \end{array} \right\rangle +2 t_2 \left\langle \begin{array}{c} \includegraphics[scale=0.35]{B2.eps} \end{array} \right\rangle = N- \left\langle \begin{array}{c} \includegraphics[scale=0.35]{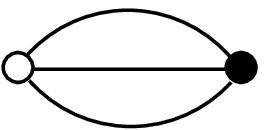} \end{array} \right\rangle.
\ee
There are two bubbles in $\mathcal{B}_{12}(2)$, but only one equation. For $p=1$, there are two independent equations on three bubbles,
\be \label{p=1}
\begin{aligned}
2t_1 \left\langle \begin{array}{c} \includegraphics[scale=0.3]{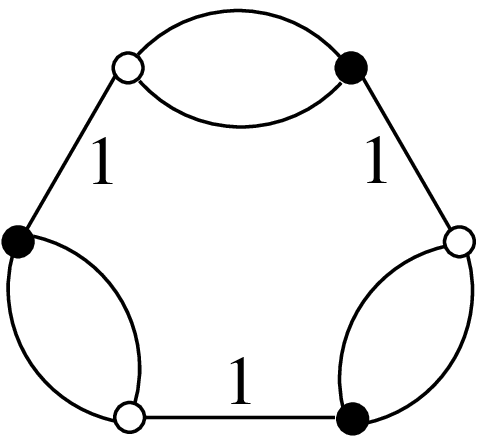} \end{array} \right\rangle +2 t_2 \left\langle \begin{array}{c} \includegraphics[scale=0.35]{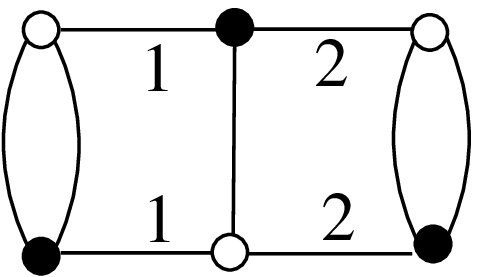} \end{array} \right\rangle &= \left\langle \begin{array}{c} \includegraphics[scale=0.35]{2vertex.eps} \end{array} \right\rangle - \left\langle \begin{array}{c} \includegraphics[scale=0.35]{B1.eps} \end{array} \right\rangle,\\
2t_1 \left\langle \begin{array}{c} \includegraphics[scale=0.35]{6vertex12.eps} \end{array} \right\rangle +2 t_2 \left\langle \begin{array}{c} \includegraphics[scale=0.3]{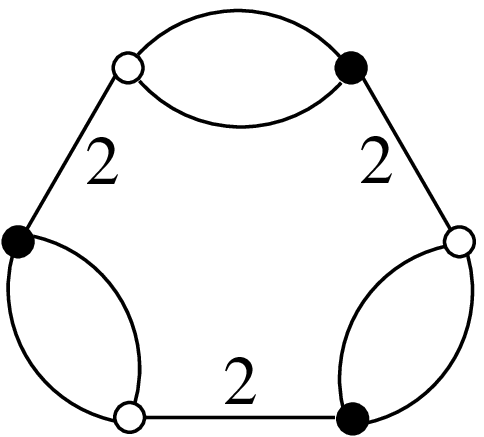} \end{array} \right\rangle &= \left\langle \begin{array}{c} \includegraphics[scale=0.35]{2vertex.eps} \end{array} \right\rangle - \left\langle \begin{array}{c} \includegraphics[scale=0.35]{B2.eps} \end{array} \right\rangle.
\end{aligned}
\ee
For $p=2$, there is again one equation less than the number of bubbles $|\mathcal{B}_{12}(4)|=6$ (the notation $1\leftrightarrow 2$ means that another equation is obtained by exchanging the colors 1 and 2, and $t_1$ with $t_2$),
\be \label{p=2}
\begin{aligned}
2t_1 \left\langle \begin{array}{c} \includegraphics[scale=0.3]{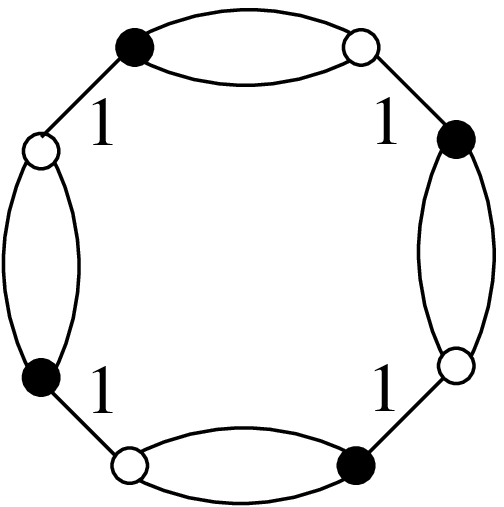} \end{array} \right\rangle +2 t_2 \left\langle \begin{array}{c} \includegraphics[scale=0.3]{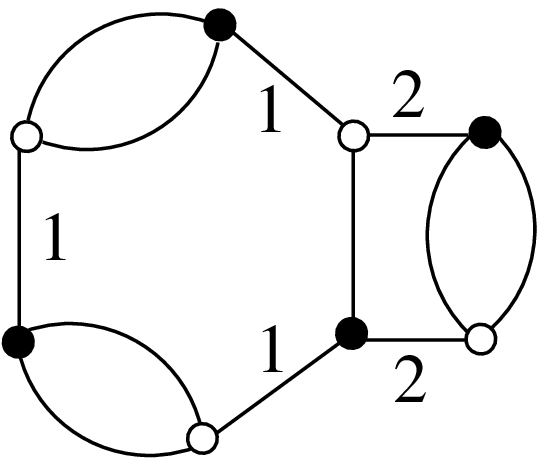} \end{array} \right\rangle &= \left\langle \begin{array}{c} \includegraphics[scale=0.35]{B1.eps} \end{array} \right\rangle - \left\langle \begin{array}{c} \includegraphics[scale=0.3]{6vertex1.eps} \end{array} \right\rangle \qquad \text{and $1\leftrightarrow2$},\\
2t_1 \left\langle \begin{array}{c} \includegraphics[scale=0.3]{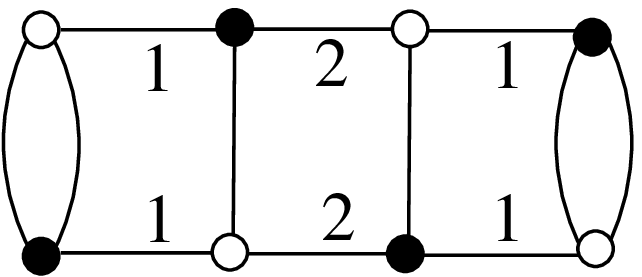} \end{array} \right\rangle +2 t_2 \left\langle \begin{array}{c} \includegraphics[scale=0.3]{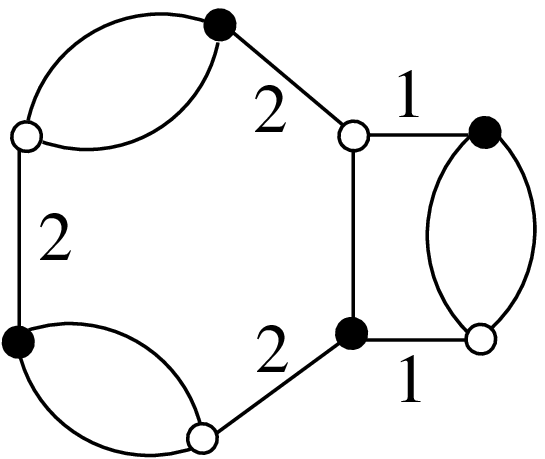} \end{array} \right\rangle &= \left\langle \begin{array}{c} \includegraphics[scale=0.35]{B1.eps} \end{array} \right\rangle - \left\langle \begin{array}{c} \includegraphics[scale=0.35]{6vertex12.eps} \end{array} \right\rangle \qquad \text{and $1\leftrightarrow2$},\\
2t_1 \left\langle \begin{array}{c} \includegraphics[scale=0.3]{8vertex12.eps} \end{array} \right\rangle +2 t_2 \left\langle \begin{array}{c} \includegraphics[scale=0.3]{8vertex21.eps} \end{array} \right\rangle &= \frac1{N} \left\langle \begin{array}{c} \includegraphics[scale=0.35]{2vertex.eps} \end{array} \right\rangle^2 - \left\langle \begin{array}{c} \includegraphics[scale=0.35]{6vertex12.eps} \end{array} \right\rangle.
\end{aligned}
\ee
However, when $p=3$ is reached, we get 12 equations, which is more than the number of bubbles on ten vertices $|\mathcal{B}_{12}(5)|=10$,
\be \label{p=3}
\begin{aligned}
2t_1 \left\langle \begin{array}{c} \includegraphics[scale=0.3]{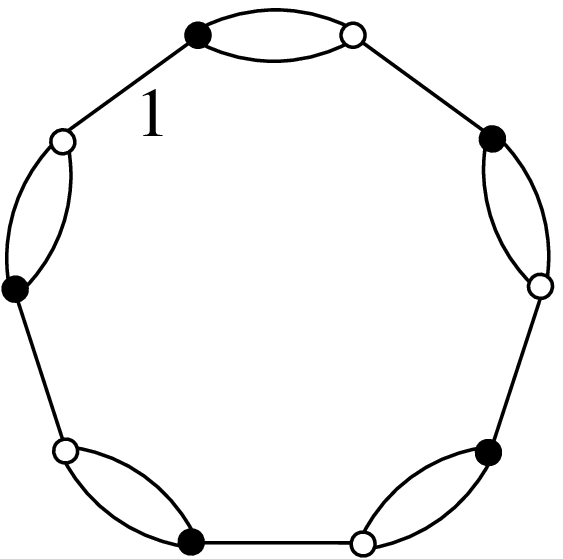} \end{array} \right\rangle +2 t_2 \left\langle \begin{array}{c} \includegraphics[scale=0.3]{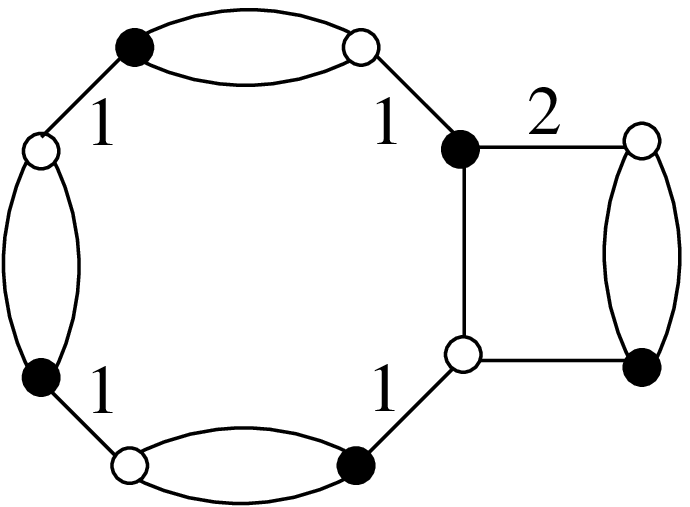} \end{array} \right\rangle &= \left\langle \begin{array}{c} \includegraphics[scale=0.3]{6vertex1.eps} \end{array} \right\rangle - \left\langle \begin{array}{c} \includegraphics[scale=0.3]{8vertex1.eps} \end{array} \right\rangle \qquad \text{and $1\leftrightarrow2$},\\
2t_1 \left\langle \begin{array}{c} \includegraphics[scale=0.3]{10vertex3.eps} \end{array} \right\rangle +2 t_2 \left\langle \begin{array}{c} \includegraphics[scale=0.3]{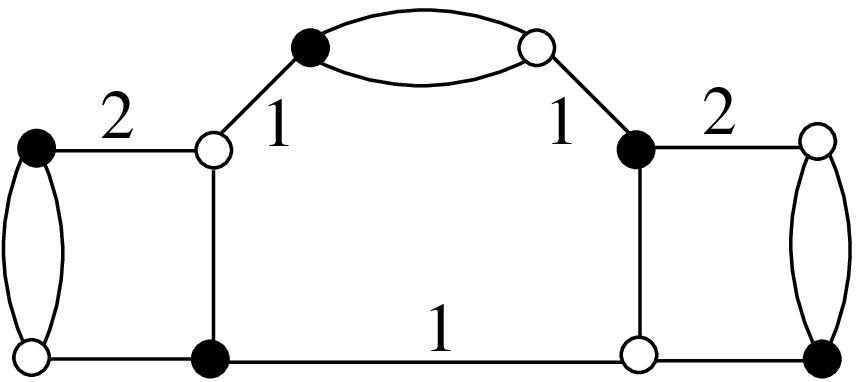} \end{array} \right\rangle &= \left\langle \begin{array}{c} \includegraphics[scale=0.3]{6vertex12.eps} \end{array} \right\rangle - \left\langle \begin{array}{c} \includegraphics[scale=0.3]{8vertex12.eps} \end{array} \right\rangle \qquad \text{and $1\leftrightarrow2$},\\
2t_1 \left\langle \begin{array}{c} \includegraphics[scale=0.35]{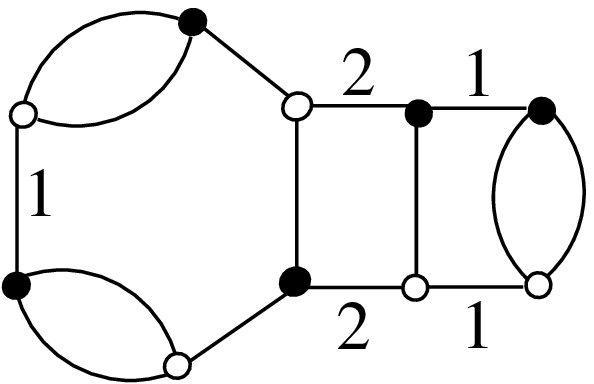} \end{array} \right\rangle +2 t_2 \left\langle \begin{array}{c} \includegraphics[scale=0.35]{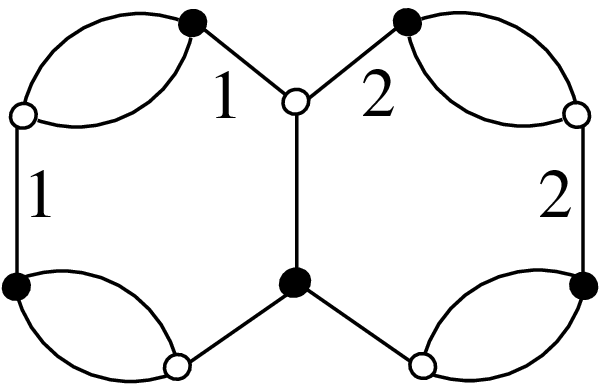} \end{array} \right\rangle &= \left\langle \begin{array}{c} \includegraphics[scale=0.3]{6vertex1.eps} \end{array} \right\rangle - \left\langle \begin{array}{c} \includegraphics[scale=0.3]{8vertex12.eps} \end{array} \right\rangle \qquad \text{and $1\leftrightarrow2$},\\
2t_1 \left\langle \begin{array}{c} \includegraphics[scale=0.3]{10vertex3.eps} \end{array} \right\rangle +2 t_2 \left\langle \begin{array}{c} \includegraphics[scale=0.35]{10vertex9.eps} \end{array} \right\rangle &= \frac{\left\langle \begin{array}{c} \includegraphics[scale=0.3]{2vertex.eps} \end{array} \right\rangle}{N} \left\langle \begin{array}{c} \includegraphics[scale=0.3]{B1.eps} \end{array} \right\rangle - \left\langle \begin{array}{c} \includegraphics[scale=0.3]{8vertex12.eps} \end{array} \right\rangle \qquad \text{and $1\leftrightarrow2$},\\
2t_1 \left\langle \begin{array}{c} \includegraphics[scale=0.35]{10vertex7.eps} \end{array} \right\rangle +2 t_2 \left\langle \begin{array}{c} \includegraphics[scale=0.35]{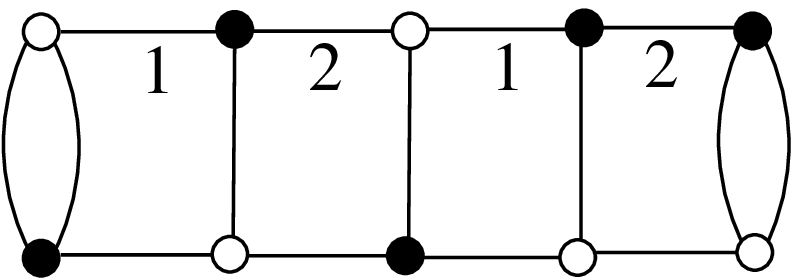} \end{array} \right\rangle &= \left\langle \begin{array}{c} \includegraphics[scale=0.35]{6vertex12.eps} \end{array} \right\rangle - \left\langle \begin{array}{c} \includegraphics[scale=0.3]{8vertex121.eps} \end{array} \right\rangle \qquad \text{and $1\leftrightarrow2$},\\
2t_1 \left\langle \begin{array}{c} \includegraphics[scale=0.35]{10vertex7.eps} \end{array} \right\rangle +2 t_2 \left\langle \begin{array}{c} \includegraphics[scale=0.3]{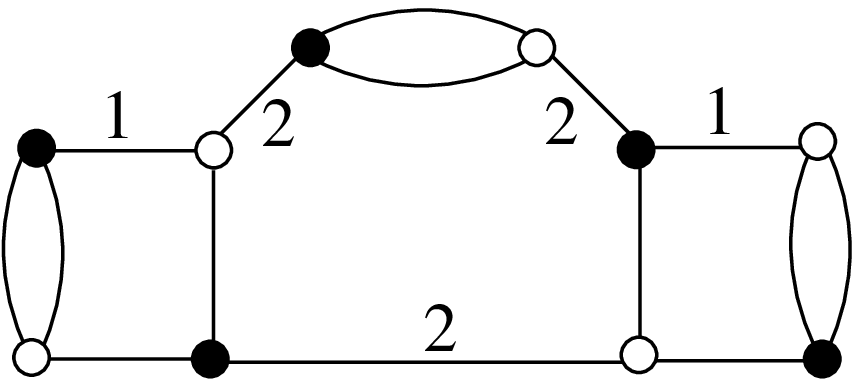} \end{array} \right\rangle &= \frac{\left\langle \begin{array}{c} \includegraphics[scale=0.3]{2vertex.eps} \end{array} \right\rangle}{N} \left\langle \begin{array}{c} \includegraphics[scale=0.3]{B1.eps} \end{array} \right\rangle - \left\langle \begin{array}{c} \includegraphics[scale=0.3]{8vertex121.eps} \end{array} \right\rangle \qquad \text{and $1\leftrightarrow2$}.
\end{aligned}
\ee
But the system is of rank $9=|\mathcal{B}_{12}(5)|-1$ only. A consequence is that it is possible to extract 3 equations which do not involve the bubbles of $\mathcal{B}_{12}(5)$ at all and appear as constraints on the BEVs of smaller bubbles. Then they might supplement the equations \eqref{p=0}, \eqref{p=1} and \eqref{p=2} and determine some of the previously left undetermined BEVs. However, this is not the case. We have checked that these constraints are trivially satisfied if \eqref{p=0}, \eqref{p=1} and \eqref{p=2} hold.

We will here refrain ourselves from writing the 35 equations for $p=4$ and the 102 equations for $p=5$ (though quite tedious, it is actually really straightforward to write them in this model since $B_1, B_2$ only add elementary melons). In those two cases, the rank is again $|\mathcal{B}_{12}(p+2)|-1$.

This inspection thus suggests that one BEV must be specified at each order in the number of vertices to determine all the others, and that the set of solutions to the SDEs is parametrized by an infinity of `initial conditions'. To see how that could work in practice, we propose to re-organize (some) SDEs according the number of melons on the color 2. %However, we show below that the SD equations can be re-organized in a way which shows that the Gaussian behavior \eqref{universality} is the only reasonable one.

When $t_2=0$, the model can reformulated as a matrix model like in the Section \ref{sec:skewed} with a simple 1-bubble potential. Then the loop observables with melons on the color 1 obey a special case of the recursion \eqref{skewed-recursion}. Therefore, it can be instructive to understand how a non-zero coupling $t_2$ affects this recursion. Denote $G_n$ the expectation value of the `loop' bubble with exactly $n$ elementary melons on the color 1. At $t_2=0$, for any $n\geq0$
\be \label{t2=0}
G_n{}_{|t_2=0} - G_{n+1}{}_{|t_2=0} -2t_1\,G_{n+2}{}_{|t_2=0} = 0.
\ee
We denote $G_{n,1}$ the expectation value of the bubble with $n$ melons on the color 1 plus one elementary melon on the color 2 (hence $n-1$ elementary melons on the color 1). This is the term generated for $t_2\neq 0$,
\be \label{G_n1}
G_n - G_{n+1} -2t_1\,G_{n+2} = 2t_2 \left\langle \begin{array}{c} \includegraphics[scale=0.35]{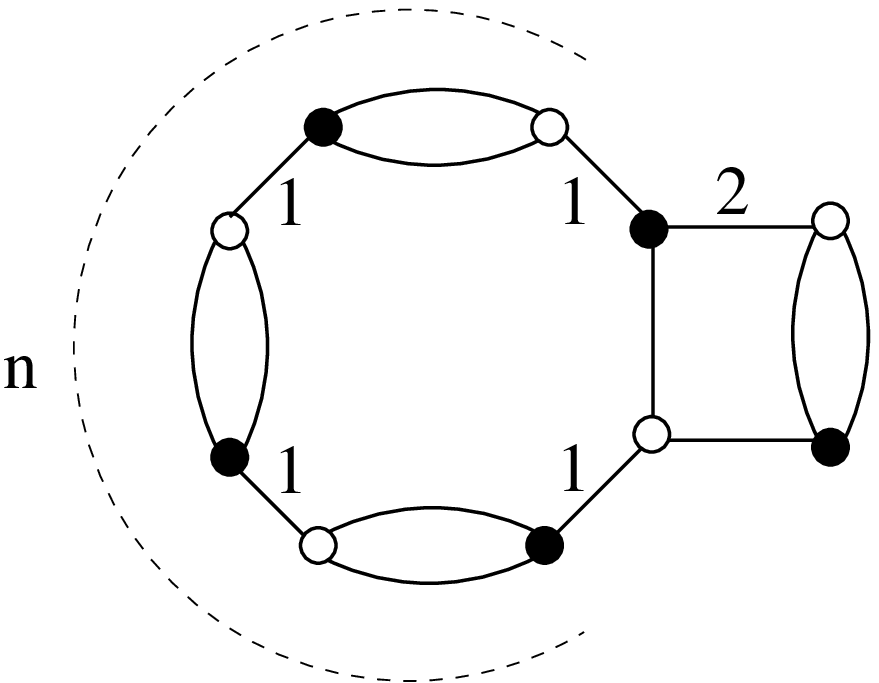} \end{array} \right\rangle
\equiv 2t_2\,G_{n+1,1}.
\ee
Let us now focus on the family $(G_{n,1})$. The bubble corresponding to $G_{n,1}$ has precisely $n-1$ elementary melons on the color 1, and another melon on the color 1 whose internal line of color 2 carries an elementary melon. By opening the graph on an elementary melon of external color 1, we see that at $t_2=0$ it satisfies the recursion \eqref{t2=0} for $n\geq1$, the reason being that the graphs simply differ by the number of elementary melons on the color 1,
\be
G_{n,1}{}_{|t_2=0} - G_{n+1,1}{}_{|t_2=0} -2t_1\,G_{n+2,1}{}_{|t_2=0} = 0.
\ee
When $t_2\neq 0$, the recursion picks up an additional term, involving graphs which have two elementary melons of external color 2. The BEVs of these graphs (for arbitrary $t_2$) only depend on the number of vertices. Indeed, this is a consequence of the Application \ref{lemma2} in the present model. The contribution of the bubble $B_1$ being the same on both sides of \eqref{eq:lemma2}, it gives the following equality\footnote{This proof only works for $t_2\neq 0$. When $t_2=0$, an independent proof is easily obtained by showing that all these graphs satisfy the same recursion, which is actually nothing but \eqref{t2=0}.}
\be \label{Gn2}
\left\langle \begin{array}{c} \includegraphics[scale=0.35]{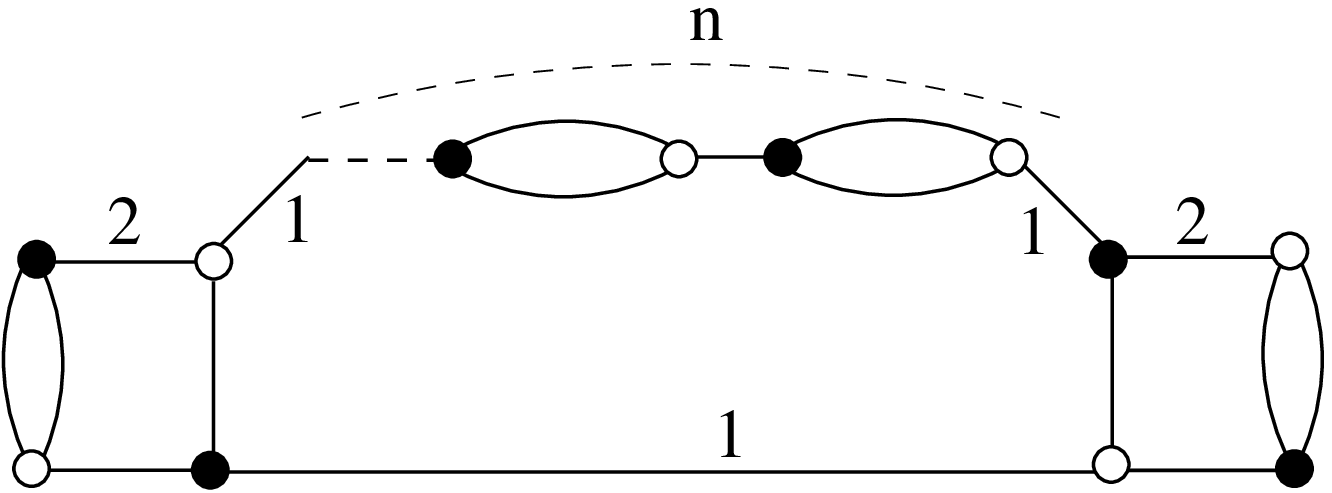} \end{array} \right\rangle = \left\langle \begin{array}{c} \includegraphics[scale=0.35]{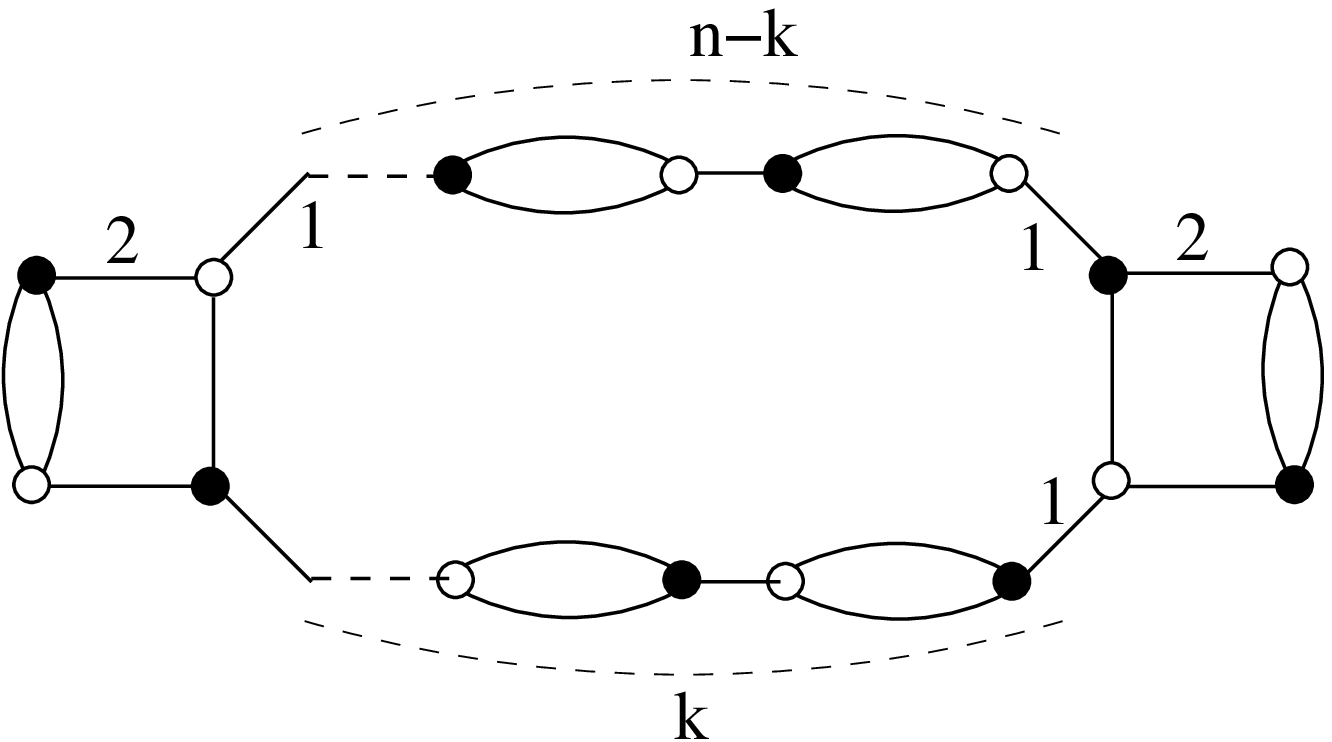} \end{array} \right\rangle.
\ee
We denote the corresponding BEV $G_{n,2}$, for $n\geq 2$ (they have exactly $n$ melons on the color 1 and two on the color 2), so that
\be \label{G_n2}
G_{n,1} - G_{n+1,1} -2t_1\,G_{n+2,1} = 2t_2\,G_{n+1,2}.
\ee

These results are similarly extended to the family of graphs $(G_{n,m})_{n\geq m}$ obtained by choosing $m$ elementary melons on the graph $G_n$ and adding an elementary melon on the color 2 on each of them. The equality \eqref{Gn2} extends to this family, stating that their BEVs are independent of the location of the $m$ elementary melons on the color 2 (as long as each of them is added on a different elementary melon with external color 1). This family satisfies a double recursion,
%These results are similarly extended to the family of graphs pictured in the Fig. ???. They possess $n$ melons of external color 1 and $m\leq n$ among them carry an elementary melon on the line of color 2 (and hence they have $n-m$ elementary melons on the color 1). Their BEVs are independent of the positions of the melons on the color 2, thanks to the Application \ref{lemma2}. Denoting them $G_{n,m}$, one gets for all $n\geq m\geq 0$ the following double-recursion,
\be
G_{n,m} - G_{n+1,m} -2t_1\,G_{n+2,m} = 2t_2\,G_{n+1,m+1}.
\ee
This double recursion does not take into account all SDEs on $\mathcal{B}_{12}$. However, it is consistent with our conjecture. It is clear it needs an infinite number of initial conditions, namely the family $(G_n)$, which can be freely chosen. It then determines the family $(G_{n,1})$ via \eqref{G_n1}, which determines in turn the family $G_{n,2}$ via \eqref{G_n2}, and so on. Again, the conjecture is then that all other SDEs either determine some BEVs outside the family $(G_{n,m})$, or are redundant.

%%%%%%%%%%%%%%%%%%
\section{Conclusion}
%%%%%%%%%%%%%%%%%%

Our main result is that the large $N$ SDEs of tensor models admit a unique physical, perturbative solution. This provides an alternative proof of the universality theorem \eqref{universality} with less combinatorics than in the original proof \cite{universality}. If we think of the SDEs as differential operators acting on the partition function and generating a specific generalization of the Virasoro algebra \cite{tree-algebra, bubble-algebra}, our result actually means that these new symmetries completely determine the physical solution. While that was a necessary step, the full representation theory of this new algebra is still to be understood. Interestingly, the representation carried by the partition function of tensor models is such that it reduces the algebra to a Virasoro algebra, as already noticed in \cite{1tensor}. 

Beyond the large $N$ limit, we expect the SDEs to give access to sub-leading orders, and possibly to a double-scaling limit (where $N$ goes to infinity and the couplings go to their critical values simultaneously). We already know that it is possible for the tensor model used in the Sec. \ref{sec:skewed} since it is actually a `very rectangular' matrix model which has been shown to possess a double-scaling limit \cite{toy-double-scaling} (and the resolvent goes like $\omega(z) = 1/(z-G) + a/N^{d-2}(z-G)^2 +b/N^{d-2}(z-G)^3+\dotsb$). This project is quite interesting since at sub-leading orders the SDEs start to distinguish bubbles that have the same number of vertices. It would be useful to find a suitable generating function for bubbles, such as the one proposed in \cite{line-colored-trees} which takes colors into account (but it is not clear whether that one is directly relevant). Another difficult part is to identify the non-melonic bubbles whose first non-zero contribution starts at a given sub-leading order.

Using those ideas, we hope that will be possible in the future to recast tensor models in a way which puts the emphasis on the new symmetries encoded by the SDEs, similarly to the fermion gas formalism of matrix models which makes the conformal field theory content clearer.

%%%%%%%%%%%%%%%%%%%%%%%%%%%%%%
\section*{Acknowledgements}

Research at Perimeter Institute is supported by the Government of Canada through Industry Canada and by the Province of Ontario through the Ministry of Research and Innovation.

\end{document}